\documentclass[journal]{IEEEtran}
\usepackage{commath,graphicx,afterpage,subfigure,amsmath,amsfonts,amssymb,algorithm,algorithmic}
\usepackage{cite,multirow,bm,multicol,booktabs,threeparttable,extpfeil}
\usepackage{amsmath,amssymb,amsfonts}
\usepackage{algorithmic}
\usepackage{graphicx}
\usepackage{algorithm,algorithmic}
\usepackage{hyperref}
\hypersetup{hidelinks=true}
\usepackage{textcomp}

\begin{document}

\title{FAConformer: Frequency-Aware Convolutional Transformer for Auditory Attention Decoding}
\author{Ziwei~Wang, Xingyi~He, Tianwang~Jia, Hongbin~Wang, and Dongrui~Wu$^{\ast}$, \IEEEmembership{Fellow,~IEEE}
\thanks{This research was supported by the National Natural Science Foundation of China (625B2077 and 62525305).}
\thanks{Z.~Wang, X.~He, T.~Jia, H.~Wang, and D.~Wu are with Hubei Key Laboratory of Brain-inspired Intelligent Systems, School of Artificial Intelligence and Automation, Huazhong University of Science and Technology, Wuhan 430074, China.}
\thanks{Ziwei Wang and Xingyi He contributed equally to this work.}
\thanks{*Corresponding Author: Dongrui Wu (drwu09@gmail.com).}}

%\markboth{IEEE Transactions on ...}
%{Shell \MakeLowercase{}}

\maketitle

\begin{abstract}
Auditory attention decoding (AAD) aims to infer the attended speaker from neural responses in multi-speaker acoustic environments and is a key problem for neuro-steered hearing systems. Although recent studies have achieved encouraging progress, existing AAD models still do not fully exploit frequency domain electroencephalography (EEG) information. In particular, most approaches introduce multi-band information through handcrafted feature extraction or direct cross-band feature concatenation, which mainly exploit frequency information at a shallow level and may overlook band-specific patterns and cross-band interactions. To address these limitations, this paper proposes FAConformer, a frequency-aware CNN-Transformer framework for AAD that explicitly integrates band-specific encoding and adaptive cross-band interaction. Specifically, FAConformer first decomposes EEG signals into multiple frequency bands and assigns each band to an independent CNN-Transformer encoder for band-specific modeling. The resulting band-wise features are then adaptively fused by a carefully designed frequency-aware attention (FAA) module that models cross-band dependencies by treating band-wise features as tokens. Further, band-wise auxiliary supervision (BAS) is introduced to prevent weakly contributing branches from being under-optimized during joint training. In this way, FAConformer performs frequency-aware modeling that more effectively exploits frequency domain information. Extensive experiments on two public AAD datasets with three decision-window lengths demonstrated that FAConformer consistently outperformed 12 competitive baselines, surpassing the current state-of-the-art model by 4.9\% and 3.0\% on DTU and KUL, respectively. Further analyses of band importance, ablation, and parameter sensitivity verify the effectiveness, robustness, and interpretability of the proposed framework. Code is available at \url{https://github.com/wzwvv/FAConformer}.
\end{abstract}

\begin{IEEEkeywords}
Brain-computer interface, Electroencephalography, auditory attention decoding, Transformer, convolutional neural network
\end{IEEEkeywords}

\section{Introduction}
\IEEEPARstart{A}{uditory} attention decoding (AAD) aims to determine which speaker a listener is attending to from neural recordings in multi-speaker acoustic environments \cite{han2019speaker}. This task is closely related to the well-known ``cocktail party'' problem \cite{cherry1953some}, where the human auditory system can selectively focus on a target speaker while suppressing competing sound sources. While this capability is essential for natural hearing in complex environments, it is often significantly degraded in hearing-impaired listeners \cite{mesgarani2012selective}. Consequently, AAD plays a crucial role in the development of neuro-steered hearing devices and brain-guided auditory assistive systems \cite{geirnaert2021electroencephalography}.

Previous studies have shown that auditory attention is strongly associated with neural activity, making it possible to study attentional selection through brain signals \cite{mesgarani2012selective,ding2012emergence}. Several neural recording modalities have been explored for AAD, including electroencephalography (EEG) \cite{o2015attentional,li2021biologically,ni2024dbpnet,yan2024darnet}, electrocorticography (ECoG) \cite{mesgarani2012selective}, and magnetoencephalography (MEG) \cite{ding2012neural,akram2016dynamic}. ECoG is invasive and used primarily in clinical settings, while MEG provides high spatial resolution but is less practical for daily use. EEG, by contrast, is non-invasive, offers high temporal resolution, and is most widely used due to its practicality \cite{faghihi2022neuroscience}.

Despite the progress, robust EEG-based AAD remains challenging. EEG responses associated with auditory attention are typically weak, noisy, and highly variable over time, which makes stable discriminative representation learning difficult. Previous studies have shown that attention-related neural responses are distributed across multiple frequency bands \cite{li2021biologically,ni2024dbpnet,dai2025gcanet,jiang2022detecting,fan2025seeing}, yet frequency domain information is not yet fully exploited in most models. A common pipeline is to first decompose the EEG into several frequency bands, then extract differential entropy (DE) features onto 2D topological maps, and finally concatenate these features for classification \cite{jiang2022detecting,fan2025seeing,cai2021low}. Such a strategy mainly introduces frequency information at a shallow level, may not fully support the learning of band-specific neural patterns, and can overlook the dynamic temporal structure of EEG signals \cite{yan2024darnet}. As a result, band-specific information and cross-band dependencies may not be jointly exploited. These observations suggest that there is still room for improvement in frequency-aware AAD modeling.

Recent deep AAD studies have explored CNNs \cite{cai2021low,cai2020low,vandecappelle2021eeg,su2021auditory}, CNN-Transformers \cite{ni2024dbpnet,yan2024darnet,cai2021auditory,su2022stanet,pahuja2023xanet}, graph-based models \cite{dai2025gcanet,cai2023brain,zhou2025dhgcn}, and other sequence modeling architectures \cite{faghihi2022neuroscience,jiang2022detecting,fan2025seeing,kuruvila2021extracting,cai2023bio,gall2026corticomorphic,zhang2024swim}. CNNs are effective for local spatio-temporal EEG patterns \cite{Lawhern2018EEGNet}, whereas Transformers are well suited for modeling long-range dependencies \cite{vaswani2017attention}. Nevertheless, most existing CNN-Transformer models for AAD still operate on either raw EEG signals \cite{yan2024darnet,cai2021auditory,su2022stanet,pahuja2023xanet} or on shallowly fused multi-band features \cite{ni2024dbpnet}, rather than constructing a hierarchical modeling framework over band-specific EEG signals. In particular, two issues remain underexplored. First, each frequency band should ideally be encoded by a dedicated representation learner, since different bands may contain distinct attention-related neural dynamics. Second, the final decision should not rely on direct feature concatenation, because different frequency bands are unlikely to contribute equally or independently.

\begin{figure}[h]
\centering
\subfigure[Traditional AAD with handcrafted features]{\includegraphics[width=\linewidth,clip]{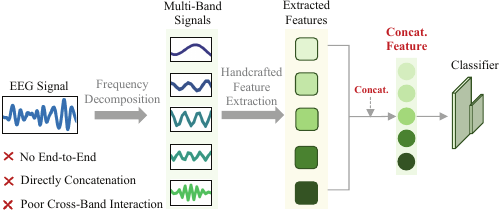}}
\subfigure[Traditional AAD with end-to-end models]{\includegraphics[width=\linewidth,clip]{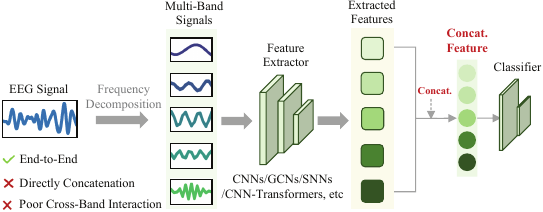}}
\subfigure[Proposed FAConformer]{\includegraphics[width=\linewidth,clip]{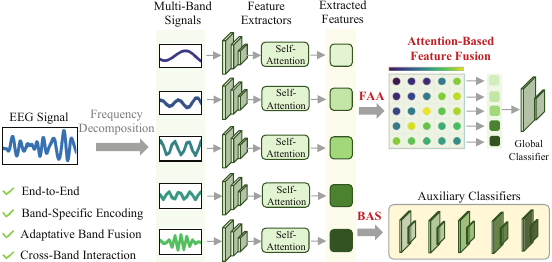}}
\caption{Comparison among (a) traditional AAD with handcrafted features, (b) traditional AAD with end-to-end models, and (c) the proposed FAConformer.}
\label{fig:intro}
\end{figure}

To address the limitations, we propose FAConformer, a frequency-aware CNN-Transformer framework for AAD. FAConformer first decomposes EEG signals into multiple frequency bands, assigning each band to an independent CNN-Transformer for band-specific encoding. The resulting band-wise features are treated as a sequence of band tokens and are adaptively fused by a frequency-aware attention (FAA) module. FAA models cross-band dependencies before final prediction. To prevent weakly contributing branches from becoming under-optimized during training, we introduce band-wise auxiliary supervision (BAS). BAS encourages each frequency branch to directly learn task-related discriminative information. Thus, FAConformer forms a complete frequency-aware hierarchical decoding pipeline, rather than simply combining multi-band decomposition and feature fusion. Fig.~\ref{fig:intro} compares FAConformer with conventional AAD models.

The main contributions of this work are.
\begin{enumerate}
\item We propose FAConformer, a frequency-aware CNN-Transformer framework for AAD that enhances within-band encoding and cross-band information fusion, thereby providing an effective way to exploit frequency domain EEG information.
\item We develop a band-specific encoding strategy, in which each band-limited EEG signal is processed by an independent CNN-Transformer encoder to learn discriminative within-band representations, jointly capturing local spatio-temporal patterns and long-range temporal dependencies.
\item We design an FAA module together with BAS. FAA treats band-wise features as a sequence of band tokens to model adaptive cross-band dependencies, while BAS prevents weakly contributing branches from being under-optimized during joint training. Together, they improve the effectiveness and reliability of hierarchical frequency-aware modeling.
\item Extensive experiments on two public AAD datasets under three decision-window lengths demonstrated that FAConformer consistently outperformed 12 competitive baselines, surpassing the current state-of-the-art model by 4.9\% and 3.0\% on DTU and KUL, respectively. Additional analyses of band importance, ablation, and parameter sensitivity further validated the effectiveness, robustness, and interpretability of the proposed framework.
\end{enumerate}

The remainder of this paper is organized as follows: Section~\ref{sect:rw} reviews related works. Section~\ref{sect:me} details the proposed FAConformer. Section~\ref{sect:er} discusses the experimental results and provides analyses. Section~\ref{sect:conclusions} draws conclusions.

\section{Related Works}\label{sect:rw}
Recent AAD models have evolved from CNN-based architectures to more diverse hybrid designs. Many studies combine CNN with sequence modeling, such as Transformer, long short-term memory (LSTM), and Mamba blocks, to capture long-range temporal dependencies in EEG signals. Other works have explored graph neural networks (GNNs) for spatial dependency modeling and spiking neural networks (SNNs) for low-power neuromorphic decoding. A brief review of representative AAD models is provided in Table~\ref{tab:aad_models}.

\begin{table*}[htbp]
\centering\setlength{\tabcolsep}{2.2mm}
\caption{A review of current AAD models.}
\renewcommand{\arraystretch}{1}
\begin{tabular}{c|cccccc}
\toprule
Model Type & Model Name & Modality & Model Structure & \# Branches & \# Bands & Dataset \\
\midrule
\multirow{5}{*}{CNN}
& CSP+CNN \cite{cai2020low} & EEG, Audio & Serial & 1 & 1 & DTU \cite{fuglsang2017noise} \\
& CNN \cite{vandecappelle2021eeg} & EEG & Serial & 1 & 1 & KUL \cite{das2016effect} \\
& CNN-CM \cite{su2021auditory} & EEG, Audio & Parallel & 2 & 1 & KUL \cite{das2016effect} \\
& SSF-CNN \cite{cai2021low} & EEG & Serial & 1 & 1 & KUL \cite{das2016effect} \\
& BIAnet \cite{li2021biologically} & EEG, Audio & Parallel & 2 & 5 & KUL \cite{das2016effect}, DTU \cite{fuglsang2017noise} \\
\midrule
\multirow{3}{*}{GNN}
& EEG-GraphNet \cite{cai2023brain} & EEG & Serial & 1 & 1 & KUL \cite{das2016effect}, DTU \cite{fuglsang2017noise} \\
& GCANet \cite{dai2025gcanet} & EEG, Audio & Parallel & 2 & 5 & KUL \cite{das2016effect}, DTU \cite{fuglsang2017noise}, AVGC \cite{rotaru2024we} \\
& DHGCN \cite{zhou2025dhgcn} & EEG & Parallel & 2 & 1 & DTU \cite{fuglsang2017noise}, MM-AAD \cite{fan2025seeing}\\
\midrule
\multirow{2}{*}{LSTM}
& CNN-LSTM \cite{kuruvila2021extracting} & EEG, Audio & Parallel & 2 & 1 & FAU \cite{kuruvila2021inference}, KUL \cite{das2016effect}, DTU \cite{fuglsang2017noise}\\
& MBSSFCC \cite{jiang2022detecting} & EEG & Serial & 1 & 5 & KUL \cite{das2016effect}, DTU \cite{fuglsang2017noise}, PKU \cite{fu2021auditory} \\
\midrule
\multirow{3}{*}{SNN}
& Faghihi \textit{et al.} \cite{faghihi2022neuroscience} & EEG & Serial & 1 & 1 & KUL \cite{das2016effect} \\
& BSAnet \cite{cai2023bio} & EEG & Serial & 1 & 1 & KUL \cite{das2016effect}, DTU \cite{fuglsang2017noise} \\
& CNN-SNN \cite{gall2026corticomorphic} & EEG, Audio & Serial & 1 & 1 & NEU \cite{Marzie2017}, DTU \cite{fuglsang2017noise} \\
\midrule
\multirow{2}{*}{Mamba}
& M-DBPNet \cite{fan2025seeing} & EEG & Parallel & 2 & 5 & KUL \cite{das2016effect}, DTU \cite{fuglsang2017noise}, MM-AAD \cite{fan2025seeing}\\
& SWIM \cite{zhang2024swim} & EEG & Serial & 1 & 1 & KUL \cite{das2016effect} \\
\midrule
\multirow{6.5}{*}{CNN-Transformer}
& STAnet \cite{su2022stanet} & EEG & Serial & 1 & 1 & KUL \cite{das2016effect}, DTU \cite{fuglsang2017noise} \\
& CMAA \cite{cai2021auditory} & EEG, Audio & Parallel & 2 & 1 & DTU \cite{fuglsang2017noise} \\
& XAnet \cite{pahuja2023xanet} & EEG & Parallel & 2 & 1 & KUL \cite{das2016effect} \\
& DBPNet \cite{ni2024dbpnet} & EEG & Parallel & 2 & 5 & KUL \cite{das2016effect}, DTU \cite{fuglsang2017noise}, MM-AAD \cite{fan2025seeing}\\
& DARNet \cite{yan2024darnet} & EEG & Serial & 1 & 1 & KUL \cite{das2016effect}, DTU \cite{fuglsang2017noise}, MM-AAD \cite{fan2025seeing}\\
\cmidrule{2-7}
& FAConformer (Ours) & EEG & Parallel & 8 & 8 & KUL \cite{das2016effect}, DTU \cite{fuglsang2017noise} \\
\bottomrule
\end{tabular}
\label{tab:aad_models}
\end{table*}

\subsection{CNNs}
CNNs are effective at learning local spatio-temporal patterns from multi-channel EEG and were among the earliest deep models adopted for AAD. Cai \textit{et al.} \cite{cai2020low} combined common spatial patterns (CSP) with CNNs and proposed CSP+CNN, where CSP was used to enhance spatial discriminability before CNN-based classification. Vandecappelle \textit{et al.} \cite{vandecappelle2021eeg} developed an end-to-end ultra-lightweight CNN with spatio-temporal convolutions for efficient AAD. CNN-CM \cite{su2021auditory} introduced a soft channel-attention mechanism to dynamically reweight EEG channels. SSF-CNN \cite{cai2021low} leveraged spectro-spatial information by extracting DE features from the alpha band, projecting features onto topological maps, and using CNNs for classification. BIAnet \cite{li2021biologically} further introduced a biologically inspired attention network to capture interactions between EEG and speech.

However, CNNs mainly rely on local receptive fields and are therefore limited in modeling long-range temporal dependencies and global feature interactions \cite{wang2025dbconformer}. Such long-context information is important for AAD, especially when attention-related neural responses evolve over time. This limitation has motivated the development of more powerful architectures that incorporate the attention mechanism.

\subsection{CNN-Transformers}
To better capture long-range temporal dependencies, recent studies have introduced Transformers into AAD. STAnet \cite{su2022stanet} used temporal and spatial attention to adaptively emphasize informative EEG channels and time samples. CMAA \cite{cai2021auditory} employed a bidirectional cross-modal attention mechanism to model interactions between EEG and speech representations. XAnet \cite{pahuja2023xanet} further exploited hemispheric asymmetry by grouping EEG channels into left and right hemispheres and modeling inter-hemispheric interactions through cross-attention.

More recent studies have explored hybrid models that combine convolutional feature extraction with attention-based sequence modeling. DBPNet \cite{ni2024dbpnet} adopted a dual-stream design to jointly encode temporal EEG dynamics along with frequency domain representations, showing the benefit of integrating temporal and spectral-spatial cues. DARNet \cite{yan2024darnet} further improved long-range modeling by coupling spatio-temporal convolutions alongside stacked Transformer encoders and dual-attention refinement. These studies indicate that CNN-Transformers are promising for AAD, especially for capturing the long-term temporal dependencies. However, present approaches still leave room for improvement in frequency-aware modeling. In particular, most works introduce frequency information via handcrafted DE features or relatively shallow frequency branches, rather than directly performing end-to-end hierarchical representation learning on raw multi-band EEG signals. Moreover, adaptive fusion across multiple frequency bands is still underexplored.

\subsection{Other Architectures}
Beyond CNNs and Transformers, several studies have explored alternative architectures for AAD, including GNNs \cite{dai2025gcanet,cai2023brain,zhou2025dhgcn}, LSTM models \cite{jiang2022detecting,kuruvila2021extracting}, SNNs \cite{faghihi2022neuroscience,cai2023bio,gall2026corticomorphic}, and Mambas \cite{fan2025seeing}.
\begin{enumerate}
\item \textit{GNNs.} GNNs aim to exploit the non-Euclidean topology of EEG channels for AAD. EEG-GraphNet \cite{cai2023brain} modeled multichannel EEG as a brain-topology graph and combined graph convolution with channel-wise attention. DHGCN \cite{zhou2025dhgcn} constructed temporal and spatial hypergraphs and performed decoding with dual hypergraph convolution. GCANet \cite{dai2025gcanet} integrated a time-frequency GNN to capture functional connectivity across brain regions and enhanced interactions between EEG and audio features by the cross-attention mechanism. GNNs improve relational modeling across EEG channels, but temporal dynamics and frequency-specific representations are not always modeled as explicitly as in sequence-based approaches.
\item \textit{LSTM models.} MBSSFCC \cite{jiang2022detecting} introduced ConvLSTM to capture spatio-temporal dependencies through gated recurrent dynamics. CNN-LSTM \cite{kuruvila2021extracting} jointly used a CNN and an LSTM, taking EEG signals and the spectrograms of multiple speakers as inputs.
\item \textit{SNNs.} Faghihi \textit{et al.} \cite{faghihi2022neuroscience} designed a three-layer SNN for AAD, two of which are integrate and fire spiking neurons. BSAnet \cite{cai2023bio} employed spiking neural mechanisms to mimic auditory neural coding. Inspired by the auditory cortex, CNN-SNN \cite{gall2026corticomorphic} combined EEG with multi-speaker speech envelopes.
\item \textit{Mambas.} M-DBPNet \cite{fan2025seeing} extended DBPNet by incorporating the Mamba block for efficient temporal modeling. SWIM \cite{zhang2024swim} was designed with a short-window CNN and a Mamba block, extracting both short-term and long-term temporal information.
\end{enumerate}

In summary, existing AAD models have achieved encouraging progress, yet several limitations exist. CNNs are effective in local spatio-temporal feature extraction but are less capable of modeling long-range dependencies, while Transformers, GNNs, LSTMs, SNNs, and Mambas improve feature learning from different perspectives. However, many multi-band AAD models still rely on hand-crafted DE features and fuse band-wise representations through direct concatenation. Such strategies may limit end-to-end band-specific representation learning and overlook adaptive cross-band interactions. Therefore, an effective end-to-end framework that integrates within-band modeling with adaptive cross-band fusion remains insufficiently explored.

\section{FAConformer} \label{sect:me}

\subsection{Overview}
\begin{figure*}[h] \centering
\includegraphics[width=\linewidth,clip]{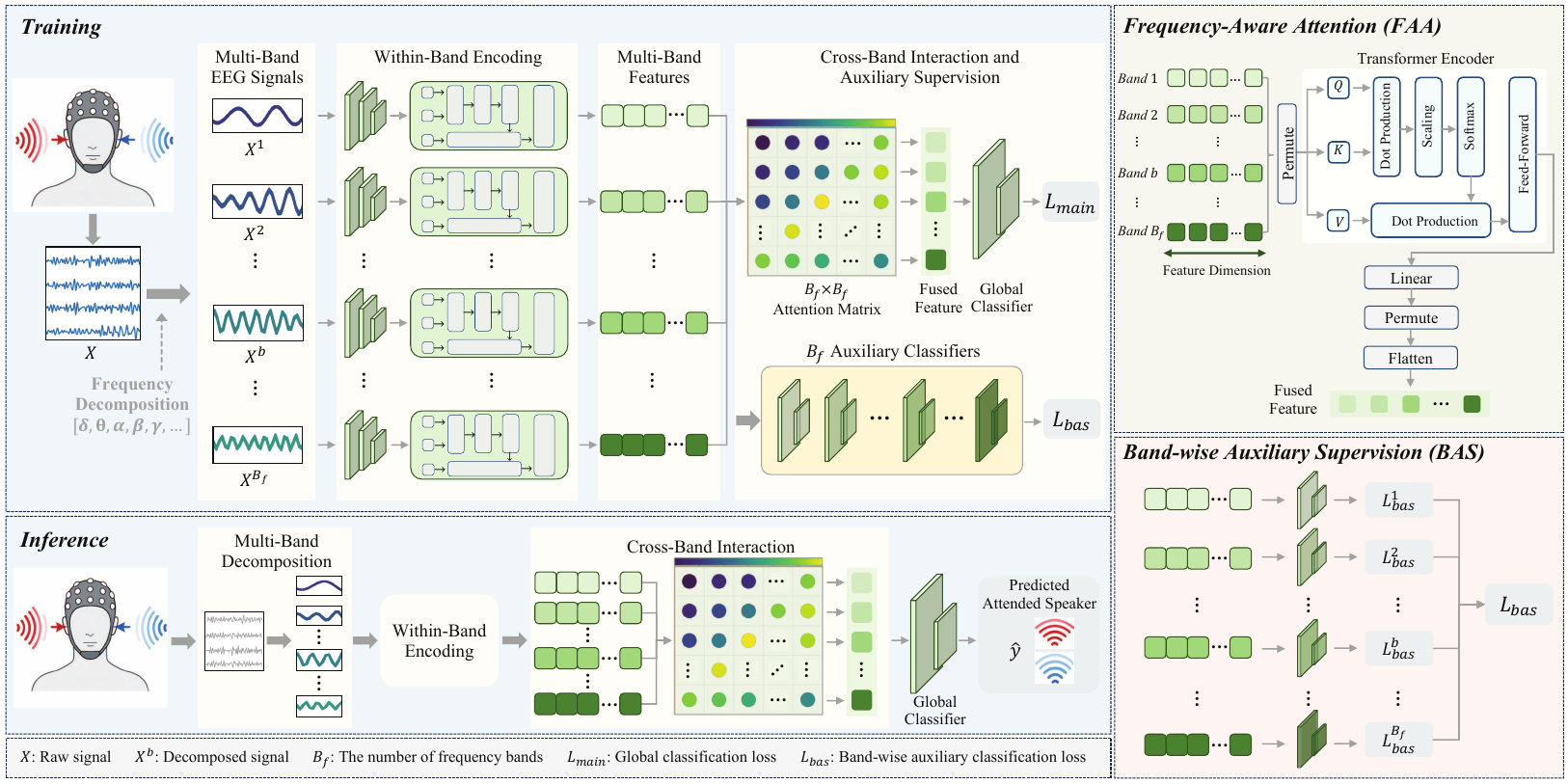}
\caption{Overall architecture of the proposed FAConformer for AAD. The upper panel shows the training framework, whereas the lower presents the inference process.
} \label{fig:framework}
\end{figure*}

This section details the proposed FAConformer, illustrated in Fig.~\ref{fig:framework}. Overall, FAConformer follows a frequency-aware hierarchical pipeline with three main components: multi-band decomposition, within-band encoding (WBE), and cross-band fusion with FAA and BAS, followed by a global classifier for final prediction.

\subsection{Network Architecture}
The detailed architecture of FAConformer is summarized in Table~\ref{tab:faconformer_network}, outlining the composition of each module. Let $\mathbf{X} \in \mathbb{R}^{C \times T}$ denote an input EEG trial, where $C$ and $T$ are the number of channels and time samples, respectively. After multi-band decomposition, $\mathbf{X}$ is transformed into a set of band-limited signals $\{\mathbf{X}^{b}\}_{b=1}^{B_f}$, where $B_f$ is the number of decomposed bands.

Specifically, each band-limited signal $\mathbf{X}^{b}$ is fed into an independent encoder to learn a band-specific representation:
\begin{equation}
\mathbf{z}^{b} = E_{b}(\mathbf{X}^{b}), \qquad b = 1,2,\ldots,B_f,
\end{equation}
where $E_{b}(\cdot)$ denotes the encoder for the $b$-th frequency band, and $\mathbf{z}^{b} \in \mathbb{R}^{D_b}$ is the corresponding feature vector.

The learned band-specific features are arranged as a sequence of band tokens and stacked along the band dimension:
\begin{equation}
\mathbf{Z} = [\mathbf{z}^{1}, \mathbf{z}^{2}, \ldots, \mathbf{z}^{B_f}] \in \mathbb{R}^{D_b \times B_f}.
\end{equation}
To capture cross-band dependencies and adaptively emphasize informative frequency components, $\mathbf{Z}$ is further fed into the proposed FAA module to perform adaptive fusion:
\begin{equation}
\mathbf{f} = F_{\mathrm{FAA}}(\mathbf{Z}),
\end{equation}
where $\mathbf{f}$ denotes the fused global representation.

Based on $\mathbf{f}$, the global classifier $C_{g}(\cdot)$ can obtain the final prediction:
\begin{equation}
\hat{y} = C_{g}(\mathbf{f}).
\end{equation}
In addition, each band-specific feature $\mathbf{z}^{b}$ is associated with the $b$-th band auxiliary classifier $C_{b}(\cdot)$:
\begin{equation}
\hat{y}^{b} = C_{b}(\mathbf{z}^{b}), \qquad b = 1,2,\ldots,B_f.
\end{equation}
The global classifier is used for the main AAD task, while the $B_f$ auxiliary classifiers are introduced to provide band-wise supervision, preventing weakly contributing branches from being under-optimized during joint training.

\begin{table*}[htpb]
\centering
\caption{FAConformer architecture.}
\centering\setlength{\tabcolsep}{1.3mm}
\renewcommand{\arraystretch}{1}
\begin{tabular}{c|c|llllll}
\toprule
Stage & Module & Layer & \# Kernels & Kernel Size & \# Parameters & Output Shape & Options \\
\midrule
\multirow{10.5}{*}{\shortstack{Within-Band \\ Encoding}}
& \multirow{7}{*}{CNNs}
& Grouped Conv1D & $F_b$ & $(1,)$ & $C \cdot F_b$ & $(B, F_b, T)$ & Channel projection \\
& & BatchNorm1D & -- & -- & $2F_b$ & $(B, F_b, T)$ & Normalization \\
& & Depthwise Conv1D & $F_b$ & $(K, K/2, \ldots)$ & -- & $(B, F_b, T)$ & Multi-scale temporal Conv. \\
& & BatchNorm1D & -- & -- & $2F_b$ & $(B, F_b, T)$ & Normalization \\
& & Aggregation, GELU & -- & -- & 0 & $(B, F_b, T)$ & Inter-path aggregation \\
& & Reshape, LogPower, Dropout & -- & $(1,W)$ & 0 & $(B, F_b, P)$ & Patch-wise embedding \\
& & Patch embeddings & -- & -- & 0 & $(B, F_b, P)$ & Patch-level feature map \\
\cmidrule{2-8}
& \multirow{3}{*}{Transformer}
& Transformer encoder & -- & -- & -- & $(B, F_b, P)$ & $L_b$ layers, $H_b$ heads \\
& & Flatten & -- & -- & 0 & $(B, F_b P)$ & Band-specific feature \\
& & Band features & -- & -- & 0 & $(B, F_b P)$ & One feature vector per band \\
\midrule
\multirow{4.5}{*}{\shortstack{Cross-Band \\ Hierarchical \\ Fusion}}
& \multirow{4}{*}{FAA}
& Stacking & -- & -- & 0 & $(B, F_b P, B_f)$ & Band-token construction \\
& & Permute & -- & -- & 0 & $(B_f, B, F_b P)$ & Sequence along band axis \\
& & Transformer encoder & -- & -- & -- & $(B_f, B, F_b P)$ & $L_f$ layers, $H_f$ heads \\
& & Linear + Flatten & -- & -- & $F_b P \cdot F_f$ & $(B, B_f F_f)$ & Adaptive frequency fusion \\
\midrule
\multirow{4.5}{*}{\shortstack{Classification \\ Heads}}
& Global & FC Layer & -- & -- & \multirow{2}{*}{$B_f F_f \rightarrow N_c$} & \multirow{2}{*}{$(B, N_c)$} & \multirow{2}{*}{Main classification head} \\
& Classifier & Norm constraint & -- & -- & & & \\
\cmidrule{2-8}
& Band-Specific & FC Layers ($\times B_f$) & -- & -- & \multirow{2}{*}{$F_bP \rightarrow N_c$} & \multirow{2}{*}{$(B, N_c)$} & \multirow{2}{*}{Auxiliary supervision heads} \\
& Classifiers & Norm constraint ($\times B_f$) & -- & -- & & & \\
\bottomrule
\end{tabular}

\vspace{0.5em}
\footnotesize{
$B$: batch size,
$B_f$: number of frequency bands,
$C$: number of EEG channels,
$T$: number of time points,
$F_b$: number of band-specific filters,
$K$: initial temporal kernel size,
$W$: patch size,
$P = T/W$: number of temporal patches,
$F_f$: FAA output dimension,
$L_b$: number of band encoder Transformer layers,
$H_b$: number of band encoder Transformer heads,
$L_f$: number of FAA Transformer layers,
$H_f$: number of FAA attention heads,
$N_c$: number of classes.
}\raggedright
\label{tab:faconformer_network}
\end{table*}

\subsection{Multi-band Decomposition}
\begin{figure}[h]\centering
\subfigure[DTU]{\includegraphics[width=.49\linewidth,clip]{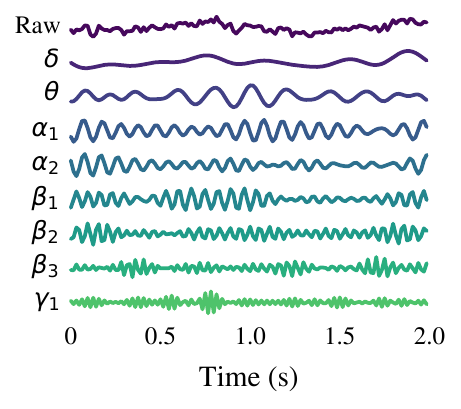}}
\subfigure[KUL]{\includegraphics[width=.49\linewidth,clip]{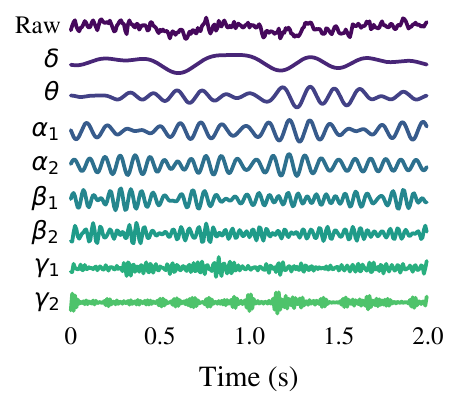}}
\caption{Visualizations of the raw EEG trial and its decomposed sub-band signals. Each sub-band highlights distinct oscillatory patterns, with temporal dynamics becoming progressively denser from low to high frequencies.} \label{fig:band_vis}
\end{figure}

To explicitly exploit frequency domain information in EEG, each trial was decomposed into $B_f=8$ canonical frequency bands for both datasets, as shown in Fig.~\ref{fig:band_vis} and Table~\ref{tab:band_info}. The band definitions were based on commonly used frequency partitioning schemes \cite{babiloni2020international}. In implementation, band decomposition was performed in the Fourier domain using the fast Fourier transform (FFT). Given an EEG trial $\mathbf{X}$, we first computed its Fourier spectrum $\mathcal{F}(\mathbf{X})$, retained the frequency bins within the $b$-th target band using a binary mask $\mathbf{M}^{b}$, and reconstructed the corresponding band-limited signal by inverse FFT:
\begin{equation}
\mathbf{X}^{b}=G_{b}(\mathbf{X})=\mathcal{F}^{-1}\left(\mathbf{M}^{b}\odot\mathcal{F}(\mathbf{X})\right),
\end{equation}
where $G_{b}$ indicates the multi-band decomposition function, and $\odot$ denotes element-wise multiplication. For real-valued EEG signals, the corresponding negative-frequency bins were also retained to preserve the real-valued time domain reconstruction. Since the DTU and KUL datasets were downsampled to different sampling rates, i.e., 64 Hz and 128 Hz, respectively, dataset-specific band boundaries were used to ensure that all frequency components remained within the valid range determined by the corresponding Nyquist frequency.

\begin{table}[htpb]
\centering\setlength{\tabcolsep}{3mm}
\renewcommand{\arraystretch}{1}
\caption{Definition of the frequency bands.}
\label{tab:band_info}
\begin{tabular}{c|ccl}
\toprule
Dataset & Band & Range (Hz) & Primary Role \\
\midrule
\multirow{8}{*}{DTU}
& $\delta$   & [1, 4)   & Slow cortical activity \\
& $\theta$   & [4, 8)   & Attention and cognition \\
& $\alpha_1$ & [8, 10)  & Early attentional modulation \\
& $\alpha_2$ & [10, 13) & Alpha suppression and control \\
& $\beta_1$  & [13, 16) & Task engagement \\
& $\beta_2$  & [16, 20) & Sustained processing \\
& $\beta_3$  & [20, 26) & Higher-order processing \\
& $\gamma_1$ & [26, 32) & Fast local integration \\
\midrule
\multirow{8}{*}{KUL}
& $\delta$   & [1, 4)   & Slow cortical activity \\
& $\theta$   & [4, 8)   & Attention and cognition \\
& $\alpha_1$ & [8, 10)  & Early attentional modulation \\
& $\alpha_2$ & [10, 13) & Alpha suppression and control \\
& $\beta_1$  & [13, 20) & Task engagement \\
& $\beta_2$  & [20, 30) & Higher-order processing \\
& $\gamma_1$ & [30, 50) & Fast local integration \\
& $\gamma_2$ & [50, 64) & High-frequency activity \\
\bottomrule
\end{tabular}
\end{table}

\subsection{Within-Band Encoding}
Each frequency band is processed by an independent band-specific encoder to learn band-specific representations. Given the $b$-th band EEG signal $\mathbf{X}^{b} \in \mathbb{R}^{C \times T}$, a series of CNN blocks are first applied to extract local spatio-temporal patterns and transform the raw signal into compact patch-level representations, and the Transformer encoder further captures long-range temporal dependencies, following recent CNN-Transformer designs in EEG decoding studies \cite{song2022eeg,zhao2025multi,liu2024msvtnet}.

Specifically, the CNN blocks, denoted by $H^{b}(\cdot)$, perform spatial projection, multi-scale temporal convolution, and patch-wise log-power embedding to generate compact patch-level features:
\begin{equation}
\mathbf{P}^{b} = H_{b}(\mathbf{X}^{b}).
\end{equation}
The resulting patch sequence is then fed into a lightweight Transformer encoder $T_{b}(\cdot)$ to model long-range dependencies within the same band and obtain the band-specific representation:
\begin{equation}
\mathbf{z}^{b} = T_{b}(\mathbf{P}^{b}), \qquad \mathbf{z}^{b} \in \mathbb{R}^{D_b}.
\end{equation}

Overall, the within-band encoding process can be formulated as
\begin{equation}
\mathbf{z}^{b} = E_{b}(\mathbf{X}^{b}),
\end{equation}
where $E_{b}(\cdot)$ denotes the encoder associated with the $b$-th band.

By assigning an independent encoder to each frequency band, FAConformer performs within-band encoding before cross-band interaction, preserving the unique characteristics of different frequency components.

\subsection{Cross-Band Hierarchical Fusion}
After within-band encoding, the band-specific feature vectors $\{\mathbf{z}^{b}\}_{b=1}^{B_f}$ are stacked along the band dimension to form a sequence of band tokens:
\begin{equation}
\mathbf{Z} = \left[\mathbf{z}^{1}, \mathbf{z}^{2}, \ldots, \mathbf{z}^{B}\right] \in \mathbb{R}^{D_b \times B_f},
\end{equation}
where each token corresponds to one frequency band.
Instead of direct concatenation, FAA organizes them into a band-token sequence for cross-band dependency modeling and adaptive fusion, as illustrated in the upper-right corner of Fig.~\ref{fig:framework}.

Specifically, the stacked band-wise features are fed into a 2-layer 2-head Transformer encoder:
\begin{equation}
\tilde{\mathbf{Z}} = T_{f}(\mathbf{Z}).
\end{equation}
By performing self-attention across the band dimension, FAA enables each frequency band to interact with all the others, thereby explicitly capturing cross-band dependencies and complementary information in a data-driven manner.

The band-wise features are then linearly projected to a target feature dimension:
\begin{equation}
\mathbf{Q} = W_{f}(\tilde{\mathbf{Z}}),
\end{equation}
where $W_{f}(\cdot)$ is a linear projection layer. Finally, the projected band-wise features are flattened into a global fused representation:
\begin{equation}
\mathbf{f} = \mathrm{Flatten}(\mathbf{Q}).
\end{equation}

Compared with direct feature concatenation, FAA has two advantages. First, it adaptively determines the contribution of each frequency band to the final decision, instead of treating all band-specific features equally. Second, it models cross-band interactions before classification, allowing the final representation to encode not only the discriminative information within each band, but also the dependency across bands.

\subsection{Band-Wise Auxiliary Supervision}
A potential issue in adaptive branch fusion is that branches assigned relatively small attention during training may not be sufficiently optimized. To address this issue, BAS is introduced, as shown in the lower-right corner of Fig.~\ref{fig:framework}.

Let $y$ denote the ground-truth label. The global classification loss is
\begin{equation}
\mathcal{L}_{\mathrm{main}}=\mathrm{CE}\!\left(\hat{y}, y\right),
\label{eq:main}
\end{equation}
where $\mathrm{CE}(\cdot,\cdot)$ is the cross-entropy loss, and $\hat{y} = C_{g}(\mathbf{f})$.

In addition to the global classifier $C_{g}(\cdot)$ applied to the fused representation $\mathbf{f}$, each band-specific feature vector $\mathbf{z}^{b}$ is associated with an auxiliary classifier $C_{b}(\cdot)$:
\begin{equation}
\hat{y}^{b} = C_b\!\left(\mathbf{z}^{b}\right), \qquad b=1,2,\ldots,B_f.
\end{equation}

For the $b$-th branch, its corresponding auxiliary loss is
\begin{equation}
\mathcal{L}^{b}_{\mathrm{bas}}=\mathrm{CE}\!\left(\hat{y}^{b}, y\right).
\end{equation}

The overall BAS loss is then computed by
\begin{equation}
\mathcal{L}_{\mathrm{bas}}=\frac{1}{B_f}\sum_{b=1}^{B_f} \mathcal{L}^{b}_{\mathrm{bas}}.
\label{eq:aux}
\end{equation}

The final training objective is
\begin{equation}
\mathcal{L}=\mathcal{L}_{\mathrm{main}}+\lambda \mathcal{L}_{\mathrm{bas}},
\label{eq:total}
\end{equation}
where $\lambda$ is a trade-off parameter balancing the global and auxiliary objectives, and is set to $1$ in this paper.

The proposed BAS serves two purposes. First, it encourages each branch encoder to learn discriminative task-related information, even if its contribution in FAA is relatively small. Second, it makes the adaptive weighting mechanism in FAA more reliable by ensuring that all branches are meaningfully optimized before fusion. Therefore, FAA and BAS are mutually complementary: FAA highlights more informative frequency branches for global decision-making, while BAS guarantees that each branch remains trainable and contributes effective band-specific representations to the final fusion.

\subsection{Inference}
During inference, only the global classifier is retained, while the band-specific auxiliary classifiers are discarded. Given an unseen EEG trial $\mathbf{X}_{test}$ from the test set $\mathcal{D}_{test}$, the band-specific features are first obtained as
\begin{equation}
\mathbf{X}_{test}^{b}=G_{b}(\mathbf{X}_{test}), \qquad
\mathbf{z}_{test}^{b}=E_{b}(\mathbf{X}_{test}^{b}).
\end{equation}
The learned band-specific features are then fused by the FAA module:
\begin{equation}
\mathbf{f}_{test}=F_{\mathrm{FAA}}\left([\mathbf{z}_{test}^{1},\mathbf{z}_{test}^{2},\ldots,\mathbf{z}_{test}^{B_f}]\right).
\end{equation}
Finally, the predicted label is obtained by
\begin{equation}
\hat{y}_{test}=C_{g}(\mathbf{f}_{test}).
\end{equation}

The pseudo-code of FAConformer is given in Algorithm \ref{alg:faconformer}.

\begin{algorithm}[h]
\caption{FAConformer}\label{alg:faconformer}
\begin{algorithmic}[1]
\REQUIRE Training set $\mathcal{D}_{train}$; test set $\mathcal{D}_{test}$; number of frequency bands $B_f$; band decomposition operators $\{G_{b}\}_{b=1}^{B_f}$; band-specific encoders $\{E_{b}\}_{b=1}^{B_f}$; FAA module $F_{\mathrm{FAA}}$; global classifier $C_{g}$; band-specific classifiers $\{C_{b}\}_{b=1}^{B_f}$; trade-off parameter $\lambda$.
\ENSURE Predicted labels $\hat{\mathbf{y}}_{test}$.
\STATE Initialize all trainable parameters
\STATE \textbf{\# Training}
\FOR{each epoch}
    \FOR{each mini-batch $\mathcal{B}=\{(\mathbf{X}_{i},y_{i})\}_{i=1}^{N}\subset\mathcal{D}_{train}$}
        \FOR{$b=1$ to $B_f$}
            \STATE $\mathbf{X}_{i}^{b}=G_{b}(\mathbf{X}_{i})$
            \STATE $\mathbf{z}_{i}^{b}=E_{b}(\mathbf{X}_{i}^{b})$
            \STATE $\hat{y}_{i}^{b}=C_{b}(\mathbf{z}_{i}^{b})$
        \ENDFOR
        \STATE $\mathbf{Z}_{i}=[\mathbf{z}_{i}^{1},\mathbf{z}_{i}^{2},\ldots,\mathbf{z}_{i}^{B_f}]$
        \STATE $\mathbf{f}_{i}=F_{\mathrm{FAA}}(\mathbf{Z}_{i})$
        \STATE $\hat{y}_{i}=C_{g}(\mathbf{f}_{i})$
        \STATE Compute $\mathcal{L}_{\mathrm{main}}$ by (\ref{eq:main})
        \STATE Compute $\mathcal{L}_{\mathrm{bas}}$ by (\ref{eq:aux})
        \STATE Compute total loss $\mathcal{L}$ by (\ref{eq:total})
        \STATE Update all trainable parameters using $\nabla \mathcal{L}$
    \ENDFOR
\ENDFOR
\STATE \textbf{\# Inference}
\FOR{each test sample $\mathbf{X}_{test}\in\mathcal{D}_{test}$}
    \FOR{$b=1$ to $B_f$}
        \STATE $\mathbf{X}_{test}^{b}=G_{b}(\mathbf{X}_{test})$
        \STATE $\mathbf{z}_{test}^{b}=E_{b}(\mathbf{X}_{test}^{b})$
    \ENDFOR
    \STATE $\mathbf{Z}_{test}=[\mathbf{z}_{test}^{1},\mathbf{z}_{test}^{2},\ldots,\mathbf{z}_{test}^{B_f}]$
    \STATE $\mathbf{f}_{test}=F_{\mathrm{FAA}}(\mathbf{Z}_{test})$
    \STATE $\hat{y}_{test}=C_{g}(\mathbf{f}_{test})$
\ENDFOR
\end{algorithmic}
\end{algorithm}

\section{Experiments and Results}\label{sect:er}
This section details the datasets, experiments, and analyses. Code for FAConformer is available at \url{https://github.com/wzwvv/FAConformer}.

\subsection{Datasets}
We conducted experiments on two representative public AAD datasets, namely  DTU \cite{fuglsang2017noise} and KUL \cite{das2016effect}. Both datasets were collected in dual-speaker listening paradigms, in which participants were instructed to focus their attention on one speaker while ignoring the other.

\begin{enumerate}
\item DTU \cite{fuglsang2017noise}: This dataset contains 64-channel EEG recordings from 18 normal-hearing subjects, sampled at 512 Hz. The task required subjects to attend to one of two simultaneously presented speakers. The two sound sources were spatially positioned at $\pm60^\circ$ relative to the listener. The speech materials were selected from Danish audiobooks narrated by male and female speakers and delivered through ER-2 earphones at 60 dB. Each subject completed 60 trials, each lasting 50 seconds.
\item KUL \cite{das2016effect}: This dataset includes 64-channel EEG recordings from 16 normal-hearing subjects, originally sampled at 8,192 Hz. Similar to DTU, each subject listened to mixtures of two competing speech streams and selectively attended to one of them. The spatialized condition was used, in which head-related transfer functions were applied to simulate sound sources at $\pm90^\circ$. Each subject completed 8 trials, each lasting about 6 minutes.
\end{enumerate}

For data preprocessing, the DTU EEG signals were high-pass filtered at 0.1 Hz, notch filtered at 50 Hz, and then downsampled to 64 Hz. The EEG signals in KUL were high-pass filtered at 0.5 Hz, downsampled to 128 Hz, and artifacts were removed following the filtering strategy in \cite{somers2018generic}. For both datasets, the resulting signals were segmented into 2s, 1s, and 0.1s trials to evaluate decoding performance under different decision-window lengths.

\begin{table*}[htpb]  \centering 
\setlength{\tabcolsep}{2.6mm}
\renewcommand{\arraystretch}{1}
\caption{Summary of the DTU and KUL datasets.}
\label{tab:dataset_info}
\begin{tabular}{c|c|c|c|c|c|c|c|c}
\toprule
\multirow{2}{*}{Dataset} & Number of & Number of & Sampling & Trial Length & Duration of & Language & Direction & \multirow{2}{*}{Task Types}\\
& Subjects & EEG Channels & Rate (Hz) & (seconds) & EEG (minutes) & of stimuli & of stimuli & \\
\midrule
DTU & 18 & 64 & 512 & 2, 1, 0.1 & 50 & Danish & $\pm$60$^{\circ}$  & \multirow{2}{*}{left/right attention track} \\
KUL & 16 & 64 & 8,192 & 2, 1, 0.1 & 48 & Dutch & $\pm$90$^{\circ}$ & \\
\bottomrule
\end{tabular}
\end{table*}

\subsection{Baseline Models}
The proposed FAConformer was compared with 12 representative baselines, which can be grouped into three categories: CNN-based, AAD-specific, and CNN-Transformer models.
\begin{enumerate}
\item EEGNet \cite{Lawhern2018EEGNet} is a lightweight CNN developed for EEG decoding. It employs a temporal convolution to capture frequency-related patterns, followed by a depthwise spatial convolution to model spatial information. A separable convolution is further adopted to improve spatio-temporal feature extraction with low computational cost.
\item SCNN \cite{deepshallow2017} is a shallow CNN motivated by the filter bank CSP \cite{Ang2008}. It uses sequential temporal and spatial convolutions to extract discriminative EEG representations in an efficient manner.
\item IFNet \cite{wang2023ifnet} performs multi-band spectral-spatial modeling by decomposing EEG trials into several predefined frequency ranges, followed by 1D spatial and temporal convolutions within each band. The resulting band-wise features are concatenated and fed into a fully connected layer for final classification.
\item DBPNet \cite{ni2024dbpnet} is an AAD-specific dual-branch architecture with temporal-frequency fusion, consisting of a temporal attention branch and a frequency residual branch.
\item DARNet \cite{yan2024darnet} is an AAD-specific dual-attention refinement network for AAD, including a spatio-temporal construction module, a dual-attention refinement module, and a feature fusion module.
\item DHGCN \cite{zhou2025dhgcn} is a dual-branch hyper GNN proposed for AAD. It incorporates a hypergraph construction module, a dual-branch hypergraph learning module, and a feature fusion module.
\item CTNet \cite{zhao2024ctnet} adopts a serial CNN-Transformer design, where a convolutional front-end first extracts local EEG patterns and a Transformer encoder is subsequently utilized to model long-range temporal dependencies.
\item TMSA-Net \cite{zhao2025tmsa} combines multi-scale convolutional feature extraction with local and global attention mechanisms to enhance EEG representation learning.
\item EEGConformer \cite{song2022eeg} consists of a convolutional block, a Transformer encoder, and a classifier. Its CNN module contains temporal and spatial convolutions followed by average pooling, and the overall architecture follows a serial CNN-Transformer pipeline.
\item MSCFormer \cite{zhao2025multi} adopts multi-branch and multi-scale CNNs with the Transformer encoder to jointly capture local and global EEG characteristics.
\item MSVTNet \cite{liu2024msvtnet} extracts local spatio-temporal representations at multiple filtered sizes and integrates an temporal branch supervision strategy to improve the learned features.
\item DBConformer \cite{wang2025dbconformer} is a dual-branch convolutional Transformer for EEG decoding. It contains a temporal branch (T-Conformer) and a spatial branch (S-Conformer), which are designed to model EEG temporal dynamics and spatial patterns, respectively.
\end{enumerate}

\subsection{Implementation}

\subsubsection{Evaluation Tasks}
To assess decoding performance across different temporal resolutions, we evaluated all models with three decision-window lengths, namely 2s, 1s, and 0.1s. For each subject, the trials were divided chronologically, with the first 90\% used as the training pool and the remaining 10\% reserved as the test set. Then, one-ninth of the training pool was randomly selected as the validation set, resulting in an overall split of 8:1:1 for training, validation, and testing. The training set was used for model optimization, the validation set for early stopping, and the test set for final performance evaluation. This protocol preserves the temporal ordering between training and test data and thus better reflects real-world deployment, where past EEG data are used to predict future attention states, similar to the chronological order setting in DBConformer \cite{wang2025dbconformer}.

\subsubsection{Evaluation Metric}
Decoding performance was evaluated using classification accuracy, which measures the proportion of EEG trials in which the attended-speaker label was correctly predicted by the model.

\subsubsection{Parameter Settings}
All experiments were repeated five times with the seed list $\{41,42,43,44,45\}$. Unless otherwise specified, all models were trained with a batch size of 32 for a maximum of 200 epochs, using early stopping with a patience of 10. The learning rate was set to $5\times10^{-4}$, and the weight decay coefficient was set to $3\times10^{-4}$ for all backbones on both datasets. The trade-off parameter $\lambda$ was set to 1 for both datasets. The number of Transformer layers $L_f$ and attention heads $H_f$ in FAA were both set to 2.

To ensure consistency with the valid frequency range after downsampling, the adopted band decomposition was adjusted according to the Nyquist frequency of each dataset. Specifically, both KUL and DTU used eight predefined frequency bands, as listed in Table~\ref{tab:band_info}. For KUL, the EEG signals were downsampled to 128 Hz, yielding a valid upper frequency limit of 64 Hz. For DTU, the EEG signals were downsampled to 64 Hz, corresponding to a Nyquist frequency of 32 Hz. Accordingly, the band boundaries were defined separately for the two datasets to ensure that all decomposed frequency components remained within the valid frequency range after downsampling.

\subsection{Main Results}
\begin{figure*}[htpb]\centering
\subfigure[DTU, 18 subjects]{\includegraphics[width=\linewidth,clip]{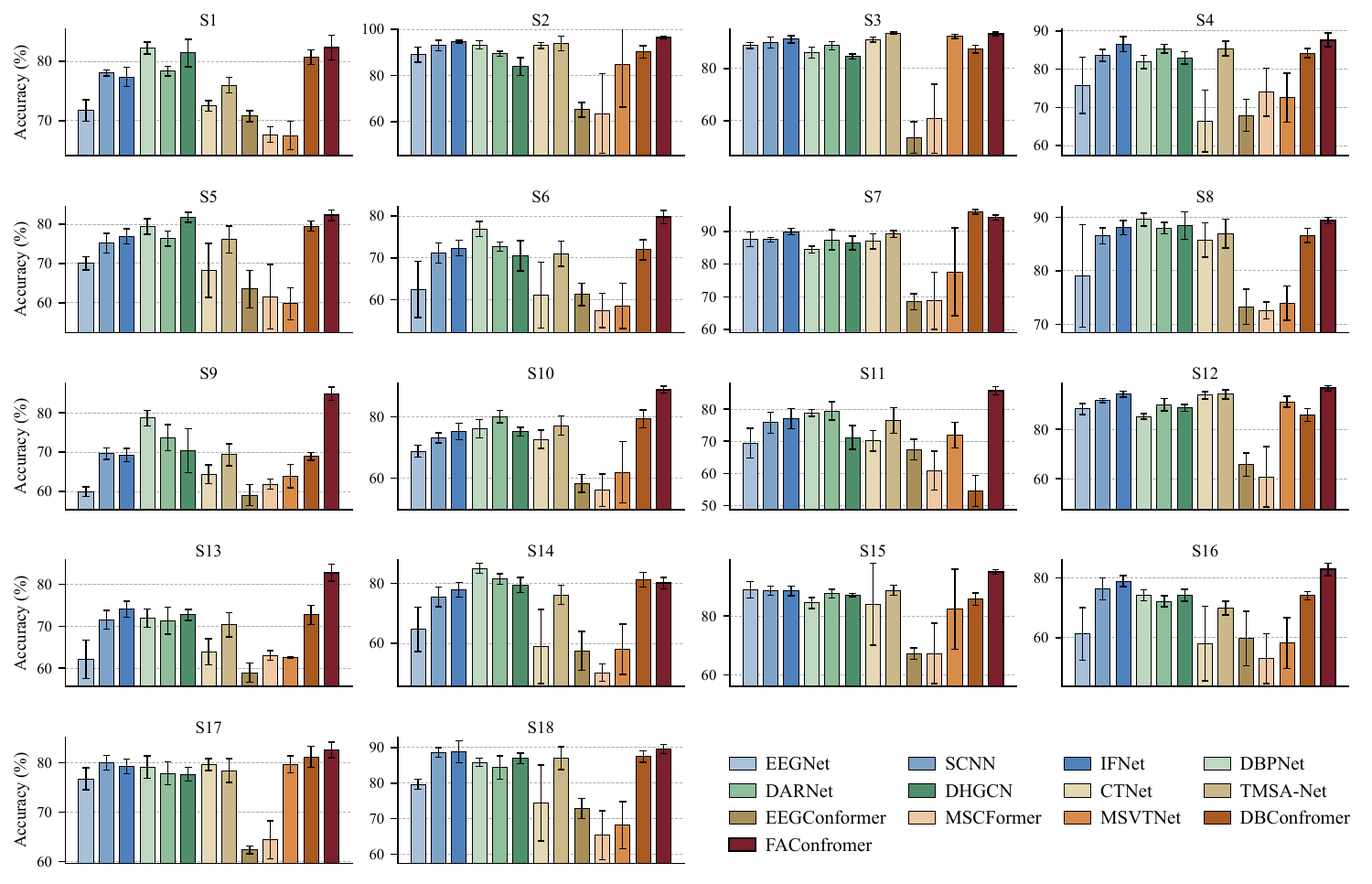}}
\subfigure[KUL, 16 subjects]{\includegraphics[width=\linewidth,clip]{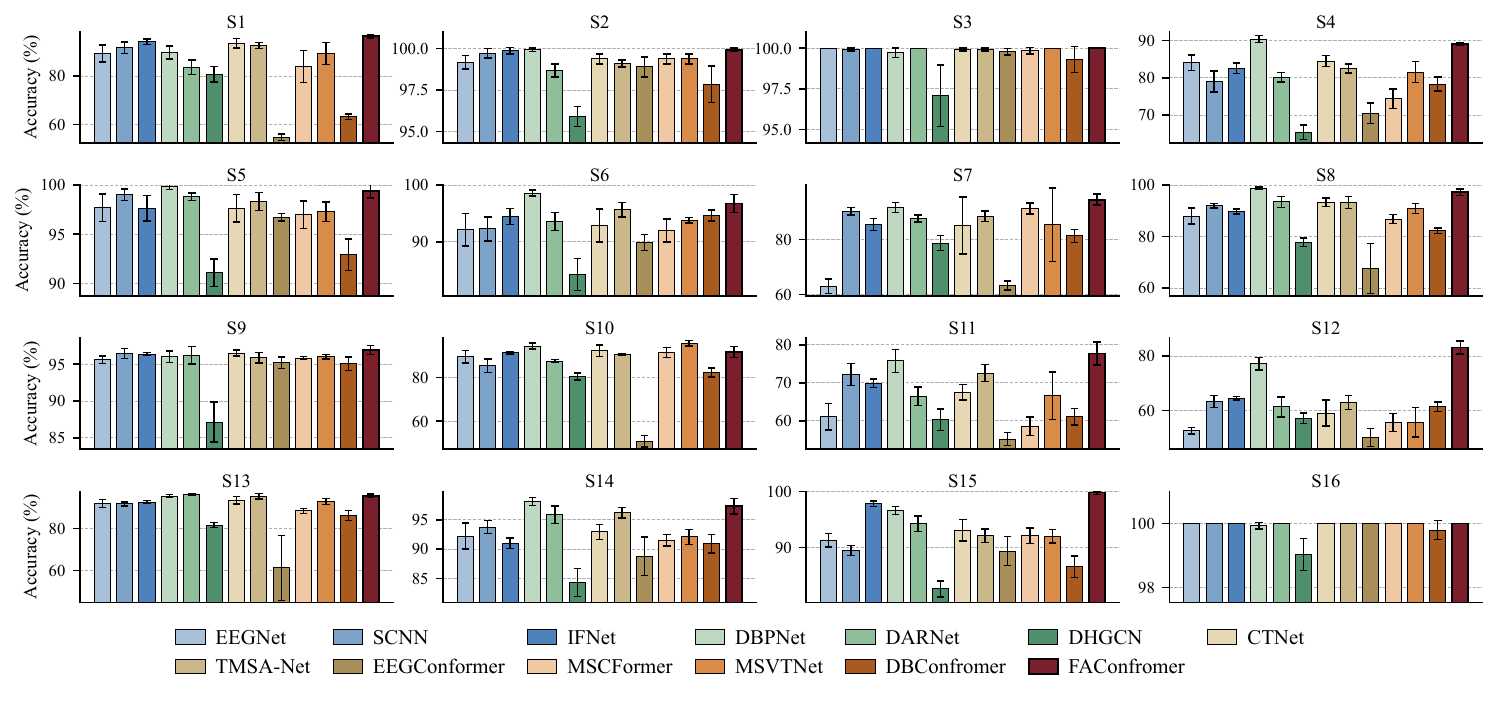}}
\caption{Subject-wise classification accuracy on (a) DTU and (b) KUL. Each subplot corresponds to one subject. Each bar denotes the mean accuracy of a model, and the error bar indicates the standard deviation.}
\label{fig:sub_acc}
\end{figure*}

The average classification results on both datasets are reported in Table~\ref{tab:main_results}, and the subject-wise classification results are illustrated in Fig.~\ref{fig:sub_acc}. The following observations can be made:
\begin{enumerate}
\item FAConformer achieved the best overall performance on both datasets. It consistently outperformed all compared baselines across decision-window lengths from 2s to 0.1s, yielding the highest average accuracy on both DTU and KUL.
\item FAConformer maintained stable superiority under different decision-window lengths. The proposed model remained consistently better than the compared baselines when the window length varied. In particular, on KUL, its advantage became more evident under shorter decision-windows, outperforming the second-best IFNet by 3.44\%. This result suggested that the frequency-aware design can provide more reliable representations when the available temporal context is limited, by jointly exploiting band-specific information and cross-band dependencies.
\item FAConformer achieved strong subject-wise performance, especially on challenging subjects. As shown in Fig.~\ref{fig:sub_acc}, FAConformer achieved the best or second-best results for almost all subjects on both datasets. Moreover, for several difficult DTU subjects, such as S9, S10, S11, S13, and S16, all baseline models achieved accuracies below 80\%, whereas FAConformer still reached above 80\%. This indicated that the proposed model not only improved the average performance, but also enhanced robustness across subjects with relatively poor baseline decoding performance.
\end{enumerate}

\begin{table*}[htpb]
\centering
\setlength{\tabcolsep}{1.7mm}
\renewcommand\arraystretch{1.2}
\caption{Average classification accuracies (\%) of FAConformer and twelve baseline models on DTU and KUL datasets. The best average performance of each network is marked in bold, and the second-best is underlined.}
\label{tab:main_results}
\begin{tabular}{c|c|ccc|c|ccc|c}
\toprule
\multirow{2.5}{*}{Model Type} & \multirow{2.5}{*}{Model} & \multicolumn{4}{c|}{DTU} & \multicolumn{4}{c}{KUL} \\
\cmidrule{3-10}
& & 2s & 1s & 0.1s & Average
& 2s & 1s & 0.1s & Average \\
\midrule
\multirow{3}{*}{CNN}
& EEGNet & 74.68$_{\pm0.93}$ & 76.97$_{\pm0.49}$ & 70.25$_{\pm0.21}$ & 73.97 & 86.72$_{\pm0.30}$ & 89.68$_{\pm0.44}$ & 86.97$_{\pm0.38}$ & 87.79 \\
& SCNN & 80.88$_{\pm0.80}$ & 79.71$_{\pm0.33}$ & 74.67$_{\pm0.14}$ & 78.42 & 89.76$_{\pm0.51}$ & 90.38$_{\pm0.22}$ & 83.83$_{\pm0.22}$ & 87.99 \\
& IFNet & \underline{82.25}$_{\pm0.75}$ & 79.21$_{\pm0.38}$ & \underline{76.73}$_{\pm0.10}$ & \underline{79.40} & 90.46$_{\pm0.25}$ & 90.88$_{\pm0.14}$ & \underline{89.03}$_{\pm0.30}$ & 90.12 \\
\midrule
\multirow{3}{*}{AAD-Specific}
& DBPNet & 81.86$_{\pm0.29}$ & 78.52$_{\pm0.30}$ & 71.99$_{\pm0.27}$ & 77.46 & \underline{93.88}$_{\pm0.36}$ & \underline{92.51}$_{\pm0.17}$ & 86.33$_{\pm0.13}$ & \underline{90.91} \\
& DARNet & 81.35$_{\pm0.47}$ & \underline{79.87}$_{\pm0.29}$ & 74.59$_{\pm0.11}$ & 78.60 & 89.60$_{\pm0.72}$ & 90.74$_{\pm0.36}$ & 87.50$_{\pm0.36}$ & 89.28 \\
& DHGCN & 80.20$_{\pm0.55}$ & 76.82$_{\pm0.73}$ & 71.49$_{\pm0.34}$ & 76.17 & 81.47$_{\pm0.15}$ & 83.96$_{\pm0.40}$ & 82.55$_{\pm0.16}$ & 82.66 \\
\midrule
\multirow{7.5}{*}{CNN-Transformer}
& CTNet & 74.72$_{\pm1.90}$ & 75.36$_{\pm1.80}$ & 72.51$_{\pm0.60}$ & 74.20 & 90.07$_{\pm0.63}$ & 90.68$_{\pm0.42}$ & 86.41$_{\pm0.20}$ & 89.05 \\
& TMSA-Net & 81.10$_{\pm0.57}$ & 79.37$_{\pm0.37}$ & 74.20$_{\pm0.32}$ & 78.22 & 90.96$_{\pm0.23}$ & 90.36$_{\pm0.31}$ & 85.83$_{\pm0.15}$ & 89.05 \\
& EEGConformer & 64.05$_{\pm1.18}$ & 67.58$_{\pm0.96}$ & 68.32$_{\pm0.43}$ & 66.65 & 77.00$_{\pm1.26}$ & 79.42$_{\pm1.25}$ & 86.48$_{\pm0.26}$ & 80.97 \\
& MSCFormer & 62.73$_{\pm1.14}$ & 67.29$_{\pm0.41}$ & 68.42$_{\pm0.47}$ & 66.15 & 87.37$_{\pm0.69}$ & 86.80$_{\pm0.35}$ & 86.06$_{\pm0.25}$ & 86.74 \\
& MSVTNet & 71.39$_{\pm2.44}$ & 73.52$_{\pm0.70}$ & 67.09$_{\pm0.30}$ & 70.67 & 89.28$_{\pm0.70}$ & 88.96$_{\pm0.32}$ & 84.93$_{\pm0.44}$ & 87.72 \\
& DBConformer & 80.42$_{\pm0.32}$ & 79.08$_{\pm0.51}$ & 76.17$_{\pm0.17}$ & 78.56 & 84.56$_{\pm0.36}$ & 85.65$_{\pm0.55}$ & 87.30$_{\pm0.21}$ & 85.84 \\
\cmidrule{2-10}
& FAConformer (Ours) & \textbf{87.48}$_{\pm0.38}$ & \textbf{84.93}$_{\pm0.52}$ & \textbf{80.72}$_{\pm0.17}$ & \textbf{84.38} & \textbf{94.71}$_{\pm0.34}$ & \textbf{94.58}$_{\pm0.17}$ & \textbf{92.47}$_{\pm0.13}$ & \textbf{93.92} \\
\bottomrule
\end{tabular}
\end{table*}

\subsection{Band Importance Analysis}
\begin{figure*}[htpb]\centering
\subfigure[DTU, 18 subjects]{\includegraphics[width=\linewidth,clip]{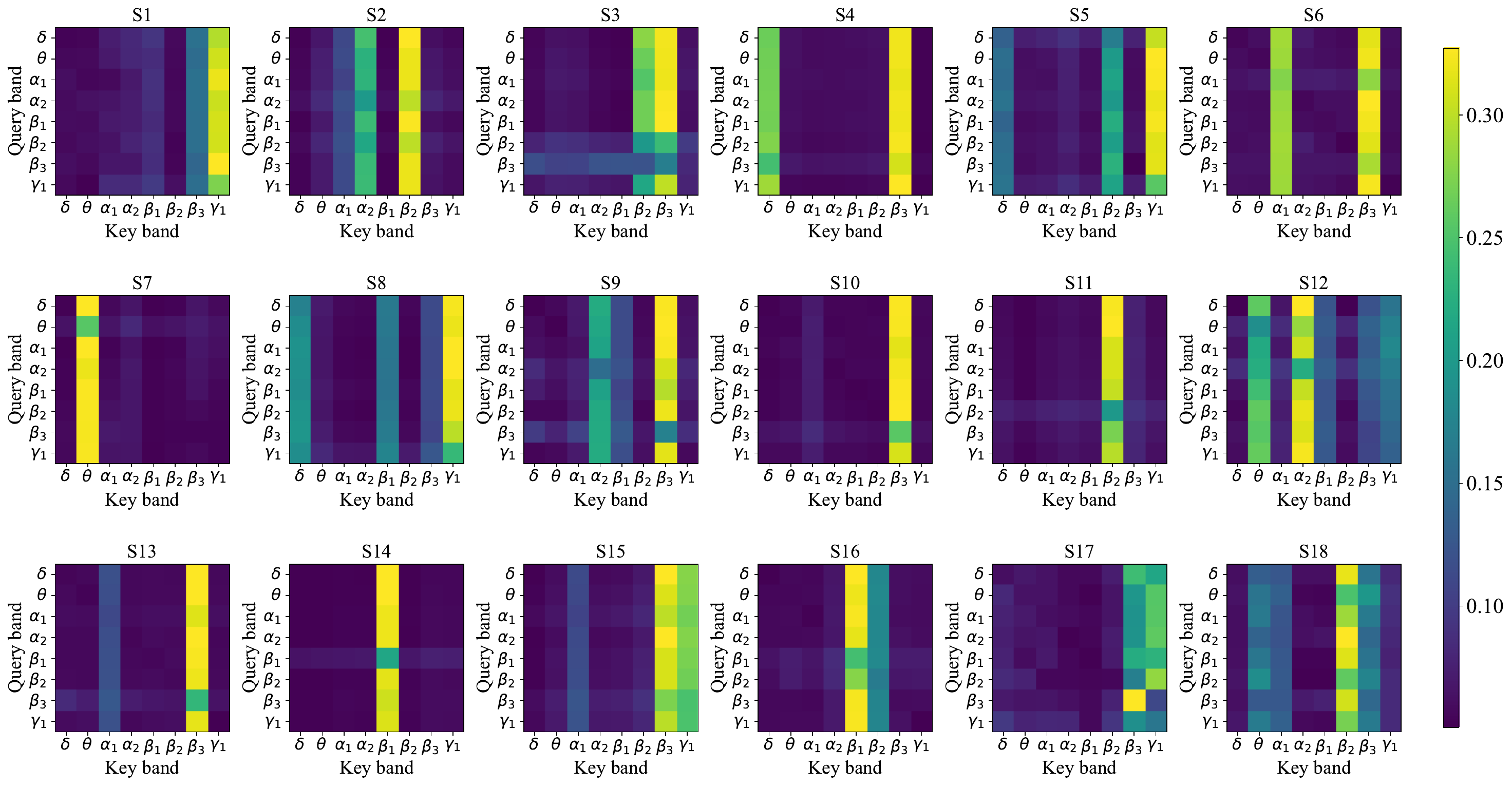}}
\subfigure[KUL, 16 subjects]{\includegraphics[width=\linewidth,clip]{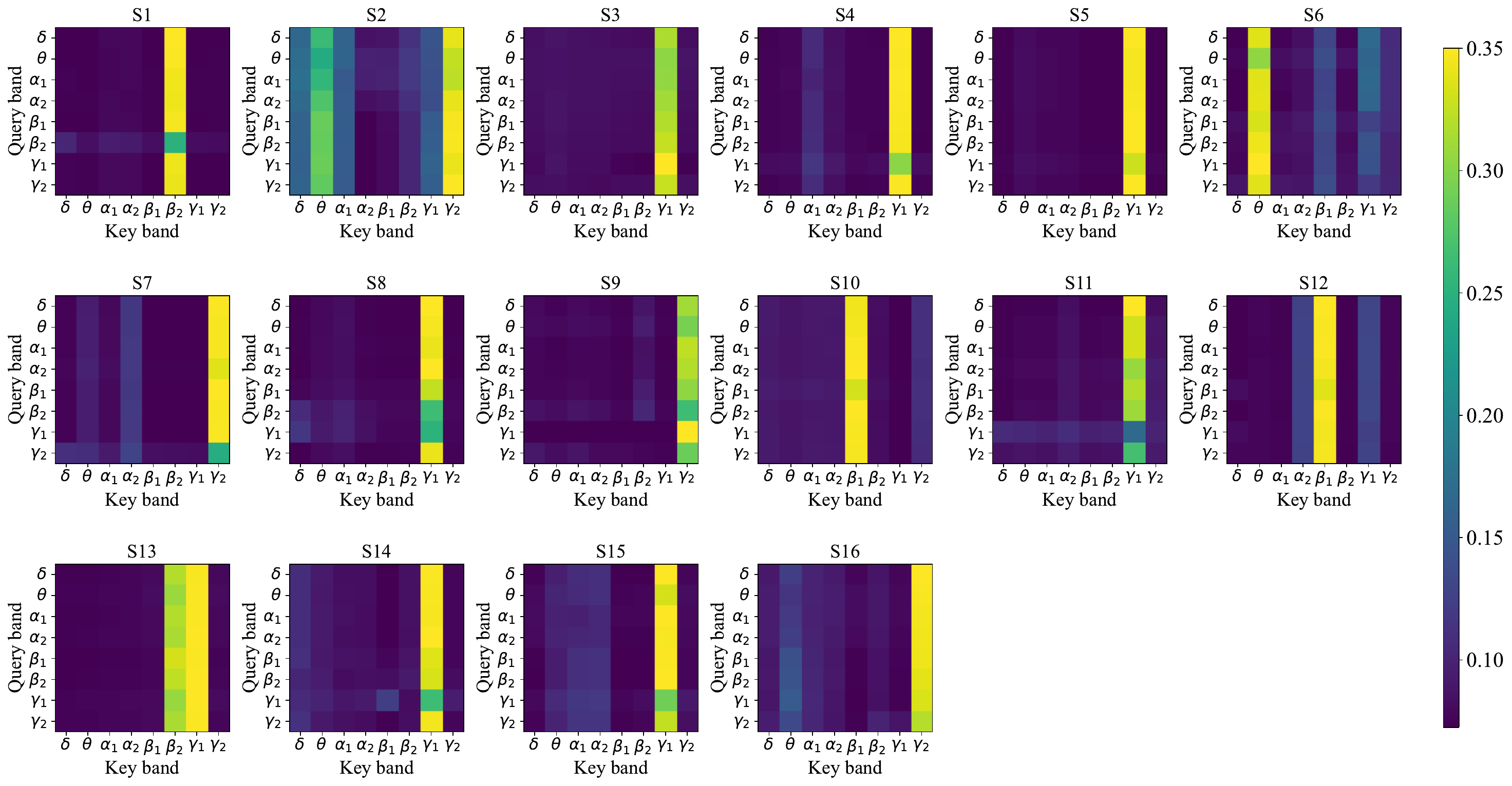}}
\caption{Subject-wise FAA cross-band attention maps on (a) DTU and (b) KUL. For each subject, the attention matrix was obtained by averaging the self-attention scores over all test samples, Transformer layers, and attention heads. Brighter values indicate stronger attention assigned from the query band to the corresponding key band.}
\label{fig:vis_attn}
\end{figure*}

To analyze how the FAA characterized band importance during cross-band fusion, we visualized the subject-wise self-attention maps learned by the proposed FAA module on both datasets. For each subject, the attention scores were averaged across all test samples, Transformer layers, and attention heads, yielding a single $B_f \times B_f$ attention matrix. The results are shown in Fig.~\ref{fig:vis_attn}. The following observations can be made:
\begin{enumerate}
\item FAA learned clearly non-uniform cross-band interaction patterns, rather than assigning similar importance to all frequency bands. On both DTU and KUL, most subjects exhibited one or a few dominant columns in the attention map, indicating that the final fused representation was typically organized around a small subset of key bands. This observation provided direct evidence that the FAA performed selective cross-band aggregation instead of simple feature averaging or uniform concatenation.
\item The learned cross-band dependency patterns were subject-dependent, but still highly structured. On DTU, the dominant bands varied across subjects, and the attention distributions were generally more diverse, with several subjects showing relatively broader interactions across adjacent beta and low-gamma bands. In contrast, the KUL dataset exhibited a more concentrated pattern, with the dominant attention more frequently assigned to the higher-frequency bands, especially $\gamma_1$ and $\gamma_2$, and a few subjects emphasizing $\beta_2$. These results suggested that FAA did not rely on a fixed fusion template shared by all subjects; instead, it adaptively selected subject-specific key bands while preserving a clear structured preference.
\item The dominant attention was often shared across multiple query bands, suggesting the existence of hub-like bands during fusion, i.e., bands that served as common key references for multiple query bands. For most subjects, the same key band received close attention from nearly all query bands, yielding a column-wise dominant pattern. This indicated that the FAA tended to organize the cross-band fusion process around a small number of informative reference bands rather than independently aggregating each band. Such a mechanism is consistent with the design motivation of FAConformer, in which within-band encoding first extracts band-specific features, and FAA subsequently performs adaptive cross-band integration based on the learned inter-band dependencies.
\end{enumerate}

\subsection{Effect of Frequency-Aware Modeling}
\begin{figure*}[htpb]\centering
\subfigure[DTU, 2s]{\includegraphics[width=\linewidth,clip]{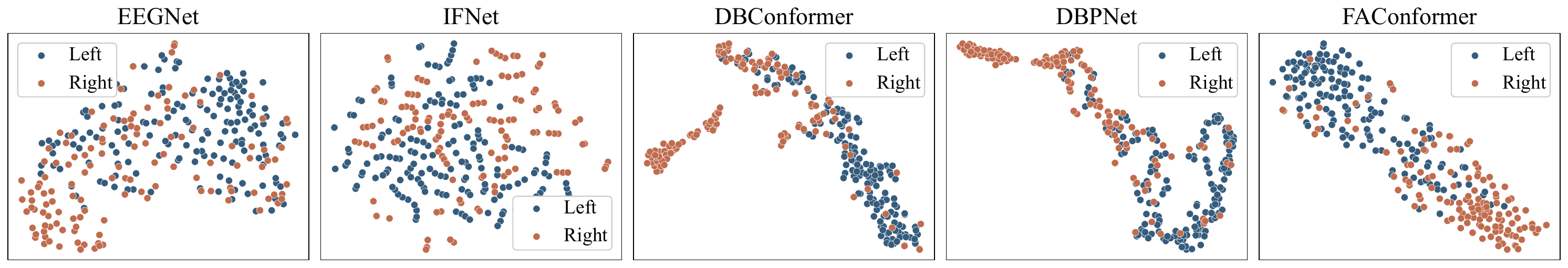}}
\subfigure[KUL, 2s]{\includegraphics[width=\linewidth,clip]{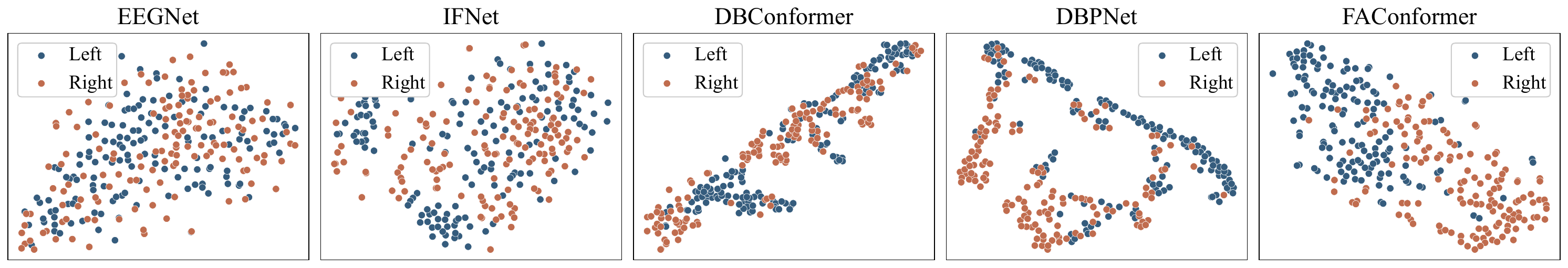}}
\subfigure[DTU, 1s]{\includegraphics[width=\linewidth,clip]{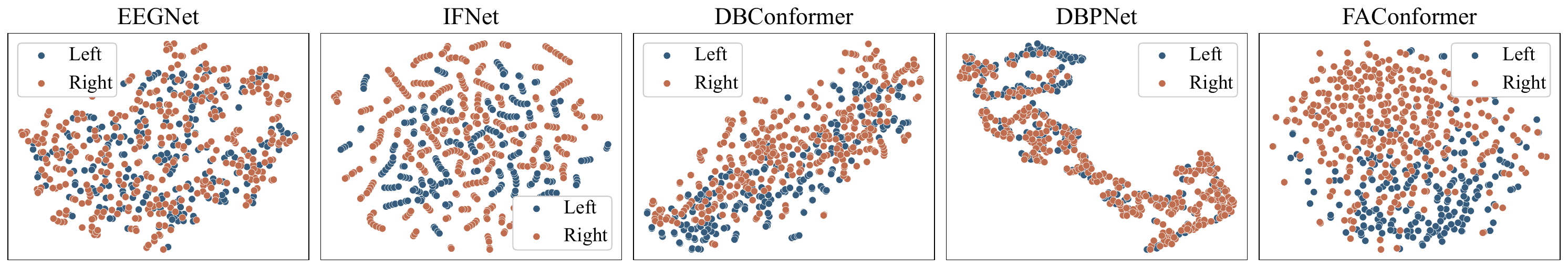}}
\subfigure[KUL, 1s]{\includegraphics[width=\linewidth,clip]{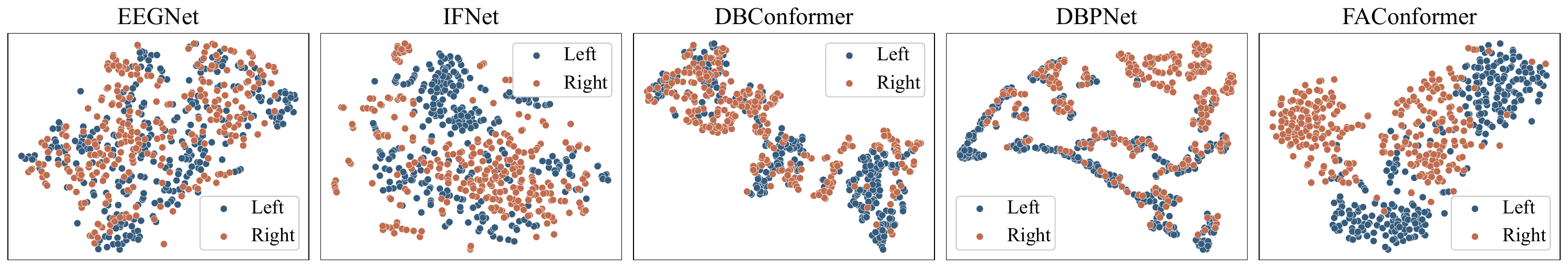}}
\subfigure[DTU, 0.1s]{\includegraphics[width=\linewidth,clip]{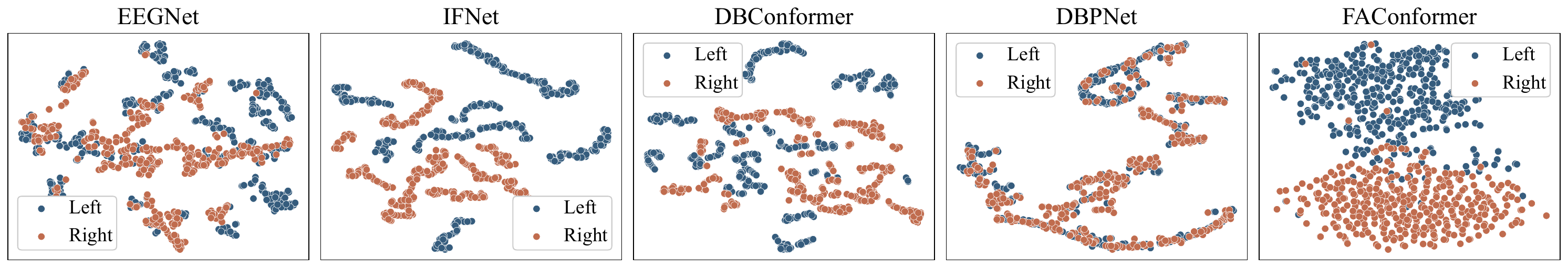}}
\subfigure[KUL, 0.1s]{\includegraphics[width=\linewidth,clip]{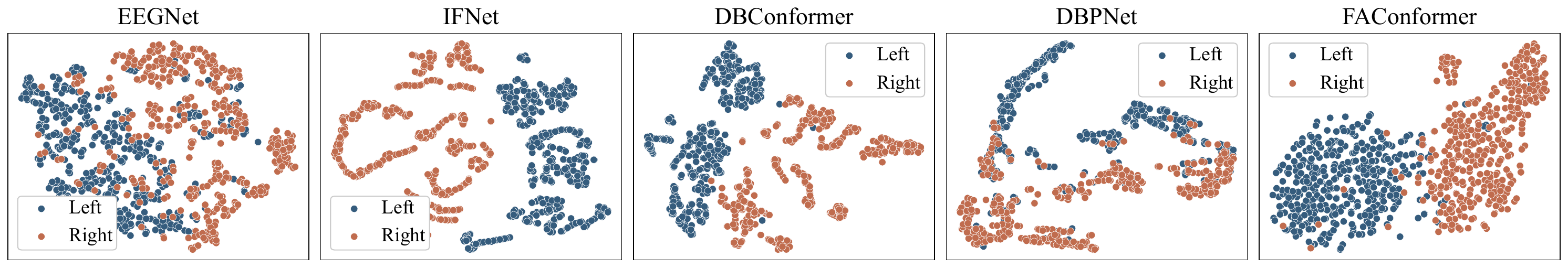}}
\caption{$t$-SNE visualizations of features extracted by EEGNet, IFNet, DBConformer, DBPNet, and FAConformer on (a) DTU with a 2s window, (b) KUL with a 2s window, (c) DTU with a 1s window, (d) KUL with a 1s window, (e) DTU with a 0.1s window, and (f) KUL with a 0.1s window. Different colors denote different attention categories.}
\label{fig:tsne}
\end{figure*}

To further investigate the effect of frequency-aware modeling on representation learning, we visualized the learned feature distributions using $t$-SNE \cite{VanderMaaten2008a}. Features extracted by EEGNet, three strong baselines (IFNet, DBConformer, and DBPNet), and the proposed FAConformer were compared on both datasets across different decision-window lengths, as shown in Fig.~\ref{fig:tsne}. Note that for the 0.1s setting, each subject contained more than 6,000 test samples due to the short window segmentation. To avoid severe overplotting, we selected the first 400 samples from each category, resulting in 800 samples in total for visualization. The following observations can be made:
\begin{enumerate}
\item FAConformer learned the most discriminative feature representations. Compared with all four baseline models, FAConformer yielded more compact intra-class clusters and clearer inter-class separation, indicating the effectiveness of the proposed frequency-aware hierarchical design in learning more structured AAD representations.
\item Among the four baseline models, DBConformer achieved relatively better feature separation. Its feature distributions showed less class overlap than those of EEGNet, IFNet, and DBPNet, consistent with its stronger quantitative performance. However, its clustering quality was still inferior to that of FAConformer.
\item The advantage of FAConformer remained consistent across different decision-window lengths. Even under the challenging 0.1s setting, FAConformer still produced comparatively well-separated feature distributions, suggesting that frequency-aware hierarchical modeling helped maintain discriminative representations when temporal context was limited.
\end{enumerate}

\subsection{Ablation Study}
The proposed frequency-aware modeling in FAConformer comprises three key components: WBE, BAS, and FAA. To clarify their individual contributions to the final performance, we conducted an ablation study by progressively adding these components to the baseline model. Table~\ref{tab:dtu_ablation} reports the ablation results on both datasets under three decision-window lengths. The following observations can be made:
\begin{enumerate}
\item WBE provided a clear and consistent performance gain. Compared with the baseline without any frequency-aware design, introducing WBE improved the accuracy from 77.25\%, 81.67\%, and 83.19\% to 78.28\%, 84.62\%, and 87.26\% on DTU under the 0.1s, 1s, and 2s settings, respectively. Similar improvements were also observed on KUL. This demonstrated that decomposing EEG into frequency bands and learning band-specific representations are beneficial for AAD.
\item BAS and FAA contributed complementary improvements. Based on WBE, adding BAS generally led to moderate improvements on both datasets, particularly under the 1s and 2s settings. This indicates that auxiliary supervision helped stabilize the optimization of band-specific branches. In contrast, adding FAA yielded larger gains under the 0.1s setting. This suggested that adaptive cross-band fusion is particularly useful when the available temporal context is limited.
\item The complete FAConformer achieved the best results in 5 of 6 settings. These results showed that WBE, FAA, and BAS jointly improved frequency-aware modeling by strengthening band-specific encoding, adaptive cross-band interaction, and reliable branch optimization.
\end{enumerate}

\begin{table}[htbp]
\centering
\setlength{\tabcolsep}{1.3mm}
\renewcommand{\arraystretch}{1.2}
\caption{Ablation study of the proposed WBE, BAS, and FAA on both datasets under three decision-window lengths.}
\label{tab:dtu_ablation}
\begin{tabular}{c|ccc|ccc}
\toprule
Dataset & WBE & BAS & FAA & 0.1s & 1s & 2s \\
\midrule
\multirow{5}{*}{DTU}
& $\times$ & $\times$ & $\times$ & 77.25$_{\pm0.24}$ & 81.67$_{\pm0.41}$ & 83.19$_{\pm0.43}$ \\
& $\checkmark$ & $\times$ & $\times$ & 78.28$_{\pm0.17}$ & 84.62$_{\pm0.36}$ & 87.26$_{\pm0.35}$ \\
& $\checkmark$ & $\checkmark$ & $\times$ & 78.45$_{\pm0.11}$ & \textbf{85.29}$_{\pm0.17}$ & \underline{87.27}$_{\pm0.40}$ \\
& $\checkmark$ & $\times$ & $\checkmark$ & \underline{80.31}$_{\pm0.25}$ & 84.06$_{\pm0.47}$ & 86.10$_{\pm0.49}$ \\
& $\checkmark$ & $\checkmark$ & $\checkmark$ & \textbf{80.72}$_{\pm0.17}$ & \underline{84.93}$_{\pm0.52}$ & \textbf{87.48}$_{\pm0.38}$ \\
\midrule
\multirow{5}{*}{KUL}
& $\times$ & $\times$ & $\times$ & 89.28$_{\pm0.29}$ & 92.29$_{\pm0.36}$ & 91.94$_{\pm0.56}$ \\
& $\checkmark$ & $\times$ & $\times$ & 91.21$_{\pm0.06}$ & 93.77$_{\pm0.08}$ & 94.06$_{\pm0.35}$ \\
& $\checkmark$ & $\checkmark$ & $\times$ & 91.33$_{\pm0.03}$ & 93.93$_{\pm0.31}$ & \underline{94.19}$_{\pm0.43}$ \\
& $\checkmark$ & $\times$ & $\checkmark$ & \underline{91.95}$_{\pm0.23}$ & \underline{94.23}$_{\pm0.31}$ & 93.92$_{\pm0.57}$ \\
& $\checkmark$ & $\checkmark$ & $\checkmark$ & \textbf{92.47}$_{\pm0.13}$ & \textbf{94.58}$_{\pm0.17}$ & \textbf{94.71}$_{\pm0.34}$ \\
\bottomrule
\end{tabular}
\end{table}

\subsection{Parameter Sensitivity Analysis}
\begin{figure}[htpb]\centering
\subfigure[$\lambda$]{\includegraphics[width=\linewidth,clip]{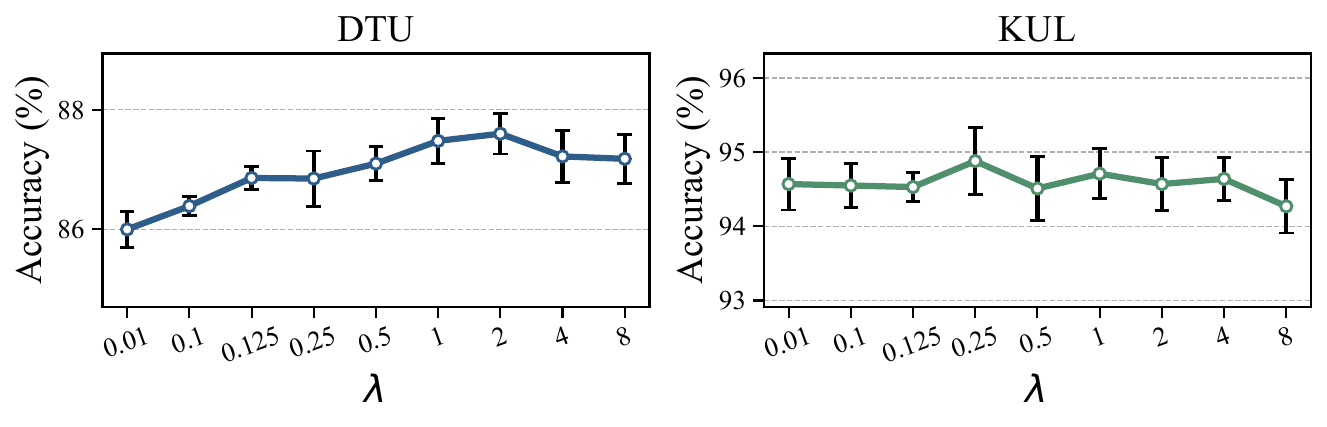}}
\subfigure[$L_f$]{\includegraphics[width=\linewidth,clip]{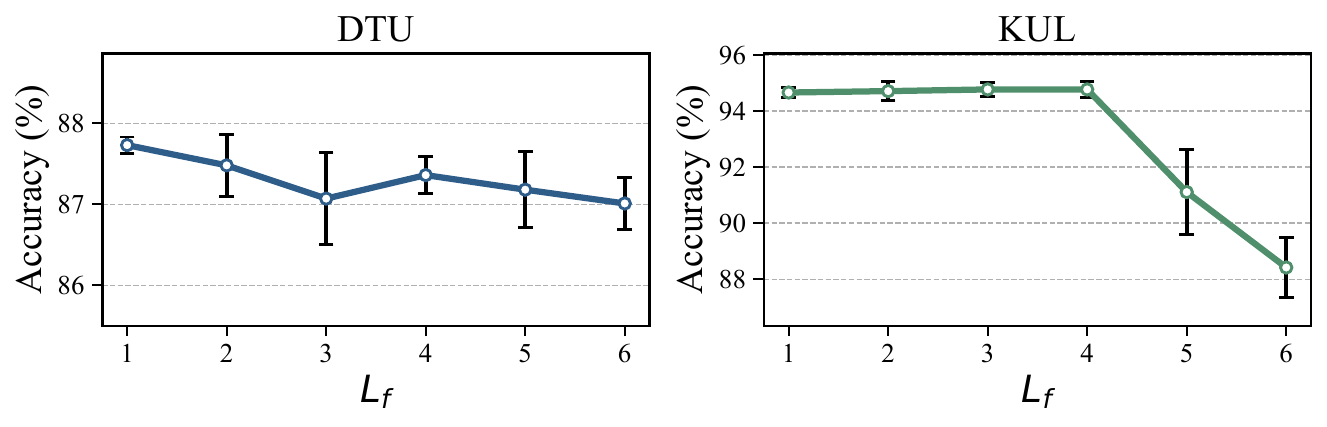}}
\subfigure[$H_f$]{\includegraphics[width=\linewidth,clip]{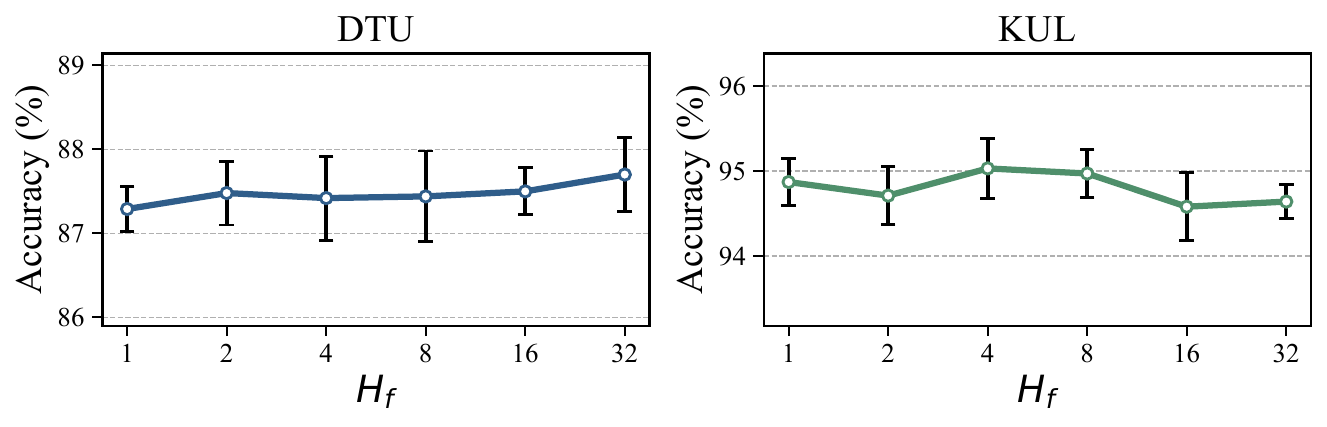}}
\caption{Parameter sensitivity analysis of FAConformer with respect to (a) $\lambda$, (b) $L_f$, and (c) $H_f$ on both datasets. The default settings of $\lambda$, $L_f$, and $H_f$ are 1, 2, and 2, respectively. Error bars denote standard deviations over repeated runs.}
\label{fig:sen_ana}
\end{figure}

To evaluate the robustness of FAConformer to hyperparameter selection, we further analyzed the sensitivity of three key parameters, namely the loss trade-off parameter $\lambda$, the number of FAA Transformer layers $L_f$, and the number of FAA Transformer heads $H_f$. The results on both datasets are shown in Fig.~\ref{fig:sen_ana}. Overall, the sensitivity analysis verified that FAConformer maintained stable performance under a wide range of hyperparameter settings, demonstrating good robustness and practical usability.
\begin{enumerate}
\item FAConformer remained stable over a wide range of $\lambda$. On DTU, the performance gradually improved as $\lambda$ increased from 0.01 to 2, and then remained relatively stable. On KUL, the accuracy varied only slightly across a broad range of $\lambda$. These results suggested that BAS made a consistently positive contribution, while the overall framework was not overly sensitive to the exact balance between the main classification loss and the auxiliary supervision loss.
\item $L_f$ had a moderate effect on DTU, but a much clearer impact on KUL. On DTU, the performance changed only slightly as $L_f$ varied from 1 to 6, indicating that a shallow FAA module was already sufficient for effective cross-band interaction. In contrast, the performance remained consistently high when $L_f \leq 4$ on KUL, but dropped noticeably when deeper FAA stacks were used. This result indicated that excessively increasing the fusion depth did not further improve the model's performance and could even weaken it.
\item FAConformer was robust to $H_f$. On DTU, the performance showed only minor fluctuations as $H_f$ increased from 1 to 32, with a slight improvement at larger head numbers. On KUL, the performance remained similarly stable. Overall, these results showed that a small number of heads was already sufficient to model effective cross-band dependencies.
\end{enumerate}

\subsection{Model Complexity Analysis}
We further compared the model complexity of all decoding models on the DTU dataset with a 2s decision-window, in terms of number of parameters, training duration, inference latency, and classification accuracy. All experiments were conducted on a single Nvidia Geforce RTX 3090 GPU. Results are summarized in Table~\ref{tab:model_complexity}. The following observations can be made:
\begin{enumerate}
\item Among the CNN baselines, IFNet provided the best accuracy-efficiency trade-off. It achieved the best performance within CNNs and the second-best overall accuracy (82.25\%) with only 8,642 parameters, indicating that it is a strong choice when computational resources are limited.
\item Among the AAD-specific models, DBPNet achieved the best accuracy, but at a relatively high model cost. Although DBPNet reached 81.86\%, it also required 915,172 parameters, which was the largest among all compared models.
\item Among the CNN-Transformer baselines, TMSA-Net and DBConformer showed relatively stronger performance, and TMSA-Net was much more lightweight.
\item FAConformer achieved the best accuracy, at the cost of increased training complexity. The proposed FAConformer obtained the highest accuracy of 87.48\%, outperforming the second-best IFNet by 5.23\%. Although FAConformer required more parameters and a longer training time, its inference latency remained within a practically acceptable range ($1.7\times10^{-2}$ s per batch). These results suggested that FAConformer offered the best performance, while its additional computational cost is acceptable in offline training and accuracy-oriented applications.
\end{enumerate}

\begin{table*}[htpb]
\centering
\setlength{\tabcolsep}{0.6mm}
\renewcommand\arraystretch{1.2}
\footnotesize
\caption{Comparison of model complexity and performance metrics across FAConformer and twelve baseline models on the DTU dataset.}
\label{tab:model_complexity}
\begin{tabular}{c|ccc|ccc|cccccc|c}
\toprule
\multirow{2.5}{*}{Metric} & \multicolumn{3}{c|}{CNN} & \multicolumn{3}{c|}{AAD-Specific} & \multicolumn{7}{c}{CNN-Transformer} \\
\cmidrule{2-14}
 & EEGNet & SCNN & IFNet & DBPNet & DARNet & DHGCN & CTNet & TMSA-Net & EEGConfor. & MSCFor. & MSVT. & DBConfor. & FAConformer \\
\midrule
\# Model Parameters & \textbf{1,066} & 103,762 & \underline{8,642} & 915,172 & 69,674 & 129,762 & 27,058 & 30,827 & 252,708 & 152,642 & 72,676 & 215,147 & 629,138 \\
Training Duration (s) & 24.51 & \textbf{12.16} & \underline{16.21} & 45.94 & 22.39 & 167.36 & 86.18 & 17.71 & 41.07 & 53.21 & 53.99 & 105.53 & 269.13 \\
Inference Latency (s) & \textbf{1.5}$e^{-3}$ & \underline{1.6}$e^{-3}$ & \textbf{1.5}$e^{-3}$ & 4.5$e^{-3}$ & 3.5$e^{-3}$ & 2.9$e^{-3}$ & 4.5$e^{-3}$ & 1.9$e^{-3}$ & 6.8$e^{-3}$ & 7.5$e^{-3}$ & 5.0$e^{-3}$ & 6.1$e^{-3}$ & 1.7$e^{-2}$ \\
Accuracy (\%) & 74.68 & 80.88 & \underline{82.25} & 81.86 & 81.35 & 80.20 & 74.72 & 81.10 & 64.05 & 62.73 & 71.39 & 80.42 & \textbf{87.48} \\
\bottomrule
\end{tabular}
\end{table*}

\section{Conclusion} \label{sect:conclusions}
This paper proposes FAConformer, a frequency-aware CNN-Transformer framework for AAD that explicitly integrates within-band encoding and adaptive cross-band interaction. Specifically, each band-limited EEG signal was filtered and processed by an independent CNN-Transformer encoder to learn band-specific representations, while the proposed FAA modeled adaptive cross-band dependencies by treating band-wise features as tokens. Further, BAS was introduced to ensure that each frequency branch remained effectively optimized during joint training. Extensive experiments on two public datasets under three decision-window lengths demonstrated the superiority of FAConformer. Additional analyses of band importance, ablation, and parameter sensitivity further validated its effectiveness, robustness, and interpretability. Future work will focus on developing more lightweight frequency-aware architectures and extending the proposed framework to more challenging AAD scenarios, including cross-subject \cite{Wang2023TASA}, cross-dataset, and multimodal settings \cite{wang2025cst}.

\bibliographystyle{IEEEtran} \bibliography{faconformer}

@InProceedings{cai2020low,
  author    = {Cai, Siqi and Su, Enze and Song, Yonghao and Xie, Longhan and Li, Haizhou},
  booktitle = {Interspeech},
  title     = {Low latency auditory attention detection with common spatial pattern analysis of {EEG} signals},
  year      = {2020},
  address   = {virtual},
  month     = {Oct.},
  pages     = {2772--2776},
}

@Article{vandecappelle2021eeg,
  author    = {Vandecappelle, Servaas and Deckers, Lucas and Das, Neetha and Ansari, Amir Hossein and Bertrand, Alexander and Francart, Tom},
  journal   = {Elife},
  title     = {{EEG}-based detection of the locus of auditory attention with convolutional neural networks},
  year      = {2021},
  pages     = {e56481},
  volume    = {10},
  publisher = {eLife Sciences Publications, Ltd},
}

@InProceedings{su2021auditory,
  author    = {Su, Enze and Cai, Siqi and Li, Peiwen and Xie, Longhan and Li, Haizhou},
  booktitle = {Int'l Conf. of the IEEE Engineering in Medicine \& Biology Society},
  title     = {Auditory attention detection with {EEG} channel attention},
  year      = {2021},
  address   = {virtual},
  month     = {Oct.},
  pages     = {5804--5807},
}

@InProceedings{cai2021low,
  author    = {Cai, Siqi and Sun, Pengcheng and Schultz, Tanja and Li, Haizhou},
  booktitle = {Int'l Conf. of the IEEE Engineering in Medicine \& Biology Society},
  title     = {Low-latency auditory spatial attention detection based on spectro-spatial features from {EEG}},
  year      = {2021},
  address   = {virtual},
  month     = {Oct.},
  pages     = {5812--5815},
}

@Article{cai2021auditory,
  author    = {Cai, Siqi and Li, Peiwen and Su, Enze and Xie, Longhan},
  journal   = {Frontiers in Neuroscience},
  title     = {Auditory attention detection via cross-modal attention},
  year      = {2021},
  pages     = {652058},
  volume    = {15},
  publisher = {Frontiers Media SA},
}

@Article{su2022stanet,
  author    = {Su, Enze and Cai, Siqi and Xie, Longhan and Li, Haizhou and Schultz, Tanja},
  journal   = {IEEE Trans. on Biomedical Engineering},
  title     = {{STA}net: A spatiotemporal attention network for decoding auditory spatial attention from {EEG}},
  year      = {2022},
  number    = {7},
  pages     = {2233--2242},
  volume    = {69},
  publisher = {IEEE},
}

@InProceedings{pahuja2023xanet,
  author    = {Pahuja, Saurav and Cai, Siqi and Schultz, Tanja and Li, Haizhou},
  booktitle = {Int'l IEEE Engineering in Medicine \& Biology Society Conf. on Neural Engineering},
  title     = {{XA}net: Cross-attention between {EEG} of left and right brain for auditory attention decoding},
  year      = {2023},
  address   = {Sydney, Australia},
  month     = {Jul.},
  pages     = {1--4},
}

@InProceedings{ni2024dbpnet,
  author    = {Ni, Qinke and Zhang, Hongyu and Fan, Cunhang and Pei, Shengbing and Zhou, Chang and Lv, Zhao},
  booktitle = {Int'l Joint Conf. on Artificial Intelligence},
  title     = {{DBPN}et: Dual-branch parallel network with temporal-frequency fusion for auditory attention detection},
  year      = {2024},
  address   = {Jeju Island, South Korea},
  month     = {Aug.},
  pages     = {3115--3123},
}

@Article{yan2024darnet,
  author  = {Yan, Sheng and Fan, Cunhang and Zhang, Hongyu and Yang, Xiaoke and Tao, Jianhua and Lv, Zhao},
  journal = {Advances in Neural Information Processing Systems},
  title   = {{DARN}et: Dual attention refinement network with spatiotemporal construction for auditory attention detection},
  year    = {2024},
  month   = {Dec.},
  pages   = {31688--31707},
  volume  = {37},
  address = {Vancouver, BC, Canada},
}

@Article{jiang2022detecting,
  author    = {Jiang, Yifan and Chen, Ning and Jin, Jing},
  journal   = {Journal of Neural Engineering},
  title     = {Detecting the locus of auditory attention based on the spectro-spatial-temporal analysis of {EEG}},
  year      = {2022},
  number    = {5},
  pages     = {056035},
  volume    = {19},
  publisher = {IOP Publishing},
}

@Article{cai2023bio,
  author    = {Cai, Siqi and Li, Peiwen and Li, Haizhou},
  journal   = {IEEE Trans. on Neural Networks and Learning Systems},
  title     = {A bio-inspired spiking attentional neural network for attentional selection in the listening brain},
  year      = {2023},
  number    = {12},
  pages     = {17387--17397},
  volume    = {35},
  publisher = {IEEE},
}

@Article{cai2023brain,
  author    = {Cai, Siqi and Schultz, Tanja and Li, Haizhou},
  journal   = {IEEE Trans. on Biomedical Engineering},
  title     = {Brain topology modeling with {EEG}-graphs for auditory spatial attention detection},
  year      = {2023},
  number    = {1},
  pages     = {171--182},
  volume    = {71},
  publisher = {IEEE},
}

@Article{fan2025seeing,
  author    = {Fan, Cunhang and Zhang, Hongyu and Ni, Qinke and Zhang, Jingjing and Tao, Jianhua and Zhou, Jian and Yi, Jiangyan and Lv, Zhao and Wu, Xiaopei},
  journal   = {Information Fusion},
  title     = {Seeing helps hearing: A multi-modal dataset and a mamba-based dual branch parallel network for auditory attention decoding},
  year      = {2025},
  pages     = {102946},
  volume    = {118},
  publisher = {Elsevier},
}

@Article{das2016effect,
  author    = {Das, Neetha and Biesmans, Wouter and Bertrand, Alexander and Francart, Tom},
  journal   = {Journal of Neural Engineering},
  title     = {The effect of head-related filtering and ear-specific decoding bias on auditory attention detection},
  year      = {2016},
  number    = {5},
  pages     = {056014},
  volume    = {13},
  publisher = {IOP Publishing},
}

@Article{fuglsang2017noise,
  author    = {Fuglsang, S{\o}ren Asp and Dau, Torsten and Hjortkj{\ae}r, Jens},
  journal   = {NeuroImage},
  title     = {Noise-robust cortical tracking of attended speech in real-world acoustic scenes},
  year      = {2017},
  pages     = {435--444},
  volume    = {156},
  publisher = {Elsevier},
}

@Article{babiloni2020international,
  author  = {Babiloni, Claudio and Barry, Robert J. and Ba{\c{s}}ar, Erol and Blinowska, Katarzyna J. and Cichocki, Andrzej and Drinkenburg, Wilhelmus HIM and Klimesch, Wolfgang and Knight, Robert T. and da Silva, Fernando Lopes and Nunez, Paul and others},
  journal = {Clinical Neurophysiology},
  title   = {International federation of clinical neurophysiology ({IFCN})-{EEG} research workgroup: Recommendations on frequency and topographic analysis of resting state {EEG} rhythms. {Part} 1: Applications in clinical research studies},
  year    = {2020},
  number  = {1},
  pages   = {285--307},
  volume  = {131},
}

@Article{kuruvila2021extracting,
  author    = {Kuruvila, Ivine and Muncke, Jan and Fischer, Eghart and Hoppe, Ulrich},
  journal   = {Frontiers in Physiology},
  title     = {Extracting the auditory attention in a dual-speaker scenario from {EEG} using a joint {CNN-LSTM} model},
  year      = {2021},
  pages     = {700655},
  volume    = {12},
  publisher = {Frontiers Media SA},
}

@Article{dai2025gcanet,
  author    = {Dai, Rui and Liao, Yuan and Han, Qiushi and Dong, Yuanlin and Yang, Yuhang and Huang, Liya},
  journal   = {Neuroscience},
  title     = {{GCAN}et: Enhancing {EEG}-based auditory attention decoding with temporal frequency {GCN} and cross attention mechanisms},
  year      = {2025},
  pages     = {212-223},
  volume    = {593},
  publisher = {Elsevier},
}

@Article{li2021biologically,
  author    = {Li, Peiwen and Cai, Siqi and Su, Enze and Xie, Longhan},
  journal   = {IEEE Signal Processing Letters},
  title     = {A biologically inspired attention network for {EEG}-based auditory attention detection},
  year      = {2021},
  pages     = {284--288},
  volume    = {29},
  publisher = {IEEE},
}

@Article{faghihi2022neuroscience,
  author    = {Faghihi, Faramarz and Cai, Siqi and Moustafa, Ahmed A.},
  journal   = {Neural Networks},
  title     = {A neuroscience-inspired spiking neural network for {EEG}-based auditory spatial attention detection},
  year      = {2022},
  pages     = {555--565},
  volume    = {152},
  publisher = {Elsevier},
}

@Article{gall2026corticomorphic,
  author  = {Gall, Richard and Kocanaogullari, Deniz and Akcakaya, Murat and Laffan, Nicole and Erdogmus, Deniz and Kubendran, Rajkumar},
  journal = {Annals of Biomedical Engineering},
  title   = {Corticomorphic hybrid {CNN-SNN} architecture for {EEG}-based low-footprint low-latency auditory attention detection},
  year    = {2026},
  pages   = {1--16},
}

@InProceedings{zhang2024swim,
  author    = {Zhang, Ziyang and Thwaites, Andrew and Woolgar, Alexandra and Moore, Brian and Zhang, Chao},
  booktitle = {2024 IEEE Spoken Language Technology Workshop},
  title     = {{SWIM}: Short-window {CNN} integrated with {Mamba} for {EEG}-based auditory spatial attention decoding},
  year      = {2024},
  address   = {Macao, China},
  month     = {Dec.},
  pages     = {1031--1038},
}

@Article{wang2025dbconformer,
  author  = {Ziwei Wang and Hongbin Wang and Tianwang Jia and Xingyi He and Siyang Li and Dongrui Wu},
  journal = {IEEE Journal of Biomedical and Health Informatics},
  title   = {{DBConformer}: Dual-branch convolutional {Transformer} for {EEG} decoding},
  year    = {2026},
  number  = {5},
  pages   = {4134--4147},
  volume  = {30},
}

@Article{song2022eeg,
  author    = {Song, Yonghao and Zheng, Qingqing and Liu, Bingchuan and Gao, Xiaorong},
  journal   = {IEEE Trans. on Neural Systems and Rehabilitation Engineering},
  title     = {{EEG} {C}onformer: Convolutional {T}ransformer for {EEG} decoding and visualization},
  year      = {2022},
  pages     = {710--719},
  volume    = {31},
  publisher = {IEEE},
}

@Article{liu2024msvtnet,
  author    = {Liu, Ke and Yang, Tao and Yu, Zhuliang and Yi, Weibo and Yu, Hong and Wang, Guoyin and Wu, Wei},
  journal   = {IEEE Journal of Biomedical and Health Informatics},
  title     = {{MSVTNet}: Multi-scale vision {T}ransformer neural network for {EEG}-based motor imagery decoding},
  year      = {2024},
  number    = {12},
  pages     = {7126-7137},
  volume    = {28},
  publisher = {IEEE},
}

@Article{somers2018generic,
  author  = {Somers, Ben and Francart, Tom and Bertrand, Alexander},
  journal = {Journal of Neural Engineering},
  title   = {A generic {EEG} artifact removal algorithm based on the multi-channel {Wiener} filter},
  year    = {2018},
  number  = {3},
  pages   = {036007},
  volume  = {15},
}

@Article{zhao2024ctnet,
  author    = {Zhao, Wei and Jiang, Xiaolu and Zhang, Baocan and Xiao, Shixiao and Weng, Sujun},
  journal   = {Scientific Reports},
  title     = {{CTNet}: A convolutional {T}ransformer network for {EEG}-based motor imagery classification},
  year      = {2024},
  number    = {1},
  pages     = {20237},
  volume    = {14},
  publisher = {Nature Publishing Group UK London},
}

@Article{Lawhern2018EEGNet,
  author  = {Lawhern, Vernon J. and Solon, Amelia J .and Waytowich, Nicholas R. and Gordon, Stephen M. and Hung, Chou P. and Lance, Brent J.},
  journal = {Journal of Neural Engineering},
  title   = {{EEGN}et: A compact convolutional neural network for {EEG}-based brain-computer interfaces},
  year    = {2018},
  number  = {5},
  pages   = {056013},
  volume  = {15},
}

@Article{deepshallow2017,
  author  = {Schirrmeister, Robin Tibor and Springenberg, Jost Tobias and Fiederer, Lukas Dominique Josef and Glasstetter, Martin and Eggensperger, Katharina and Tangermann, Michael and Hutter, Frank and Burgard, Wolfram and Ball, Tonio},
  journal = {Human Brain Mapping},
  title   = {Deep learning with convolutional neural networks for {EEG} decoding and visualization},
  year    = {2017},
  number  = {11},
  pages   = {5391--5420},
  volume  = {38},
}

@Article{wang2023ifnet,
  author    = {Wang, Jiaheng and Yao, Lin and Wang, Yueming},
  journal   = {IEEE Trans. on Neural Systems and Rehabilitation Engineering},
  title     = {{IFNet}: An interactive frequency convolutional neural network for enhancing motor imagery decoding from {EEG}},
  year      = {2023},
  pages     = {1900--1911},
  volume    = {31},
  publisher = {IEEE},
}

@Article{zhao2025tmsa,
  author    = {Zhao, Qian and Zhu, Weina},
  journal   = {Biomedical Signal Processing and Control},
  title     = {{TMSA-Net}: A novel attention mechanism for improved motor imagery {EEG} signal processing},
  year      = {2025},
  pages     = {107189},
  volume    = {102},
  publisher = {Elsevier},
}

@Article{zhao2025multi,
  author    = {Zhao, Wei and Zhang, Baocan and Zhou, Haifeng and Wei, Dezhi and Huang, Chenxi and Lan, Quan},
  journal   = {Scientific Reports},
  title     = {Multi-scale convolutional {T}ransformer network for motor imagery brain-computer interface},
  year      = {2025},
  number    = {1},
  pages     = {12935},
  volume    = {15},
  publisher = {Nature Publishing Group UK London},
}

@InProceedings{zhou2025dhgcn,
  author    = {Zhou, Jian and Xie, Yingjie and Fan, Cunhang and Wang, Huabin and Lv, Zhao and Tao, Liang},
  booktitle = {Proc. of the ACM Int'l Conf. on Multimedia},
  title     = {{DHGCN}: Dual hypergraph convolutional network for {EEG}-based auditory attention detection},
  year      = {2025},
  address   = {New York, NY, USA},
  month     = {Oct.},
  pages     = {612--620},
}

@Article{VanderMaaten2008a,
  author  = {Van der Maaten, Laurens and Hinton, Geoffrey},
  journal = {Journal of Machine Learning Research},
  title   = {{V}isualizing data using t-{SNE}},
  year    = {2008},
  number  = {11},
  pages   = {2579-2605},
  volume  = {9},
}

@InProceedings{vaswani2017attention,
  author    = {Vaswani, Ashish and Shazeer, Noam and Parmar, Niki and Uszkoreit, Jakob and Jones, Llion and Gomez, Aidan N. and Kaiser, {\L}ukasz and Polosukhin, Illia},
  booktitle = {Proc. Advances in Neural Information Processing Systems},
  title     = {Attention is all you need},
  year      = {2017},
  address   = {Long Beach, CA, USA},
  month     = {Dec.},
}

@InProceedings{Ang2008,
  author    = {Ang, Kai Keng and Chin, Zheng Yang and Zhang, Haihong and Guan, Cuntai},
  booktitle = {Proc. IEEE Int'l Joint Conf. on Neural Networks},
  title     = {Filter bank common spatial pattern ({FBCSP}) in brain-computer interface},
  year      = {2008},
  address   = {Hong Kong, China},
  month     = {Jun.},
  pages     = {2390--2397},
}

@Article{han2019speaker,
  author  = {Han, Cong and O’Sullivan, James and Luo, Yi and Herrero, Jose and Mehta, Ashesh D. and Mesgarani, Nima},
  journal = {Science Advances},
  title   = {Speaker-independent auditory attention decoding without access to clean speech sources},
  year    = {2019},
  number  = {5},
  pages   = {eaav6134},
  volume  = {5},
}

@Article{mesgarani2012selective,
  author  = {Mesgarani, Nima and Chang, Edward F.},
  journal = {Nature},
  title   = {Selective cortical representation of attended speaker in multi-talker speech perception},
  year    = {2012},
  number  = {7397},
  pages   = {233--236},
  volume  = {485},
}

@Article{cherry1953some,
  author  = {Cherry, Edward Collin},
  journal = {Journal of the Acoustical Society of America},
  title   = {Some experiments on the recognition of speech, with one and with two ears},
  year    = {1953},
  pages   = {975--979},
  volume  = {25},
}

@Article{ding2012neural,
  author    = {Ding, Nai and Simon, Jonathan Z.},
  journal   = {Journal of Neurophysiology},
  title     = {Neural coding of continuous speech in auditory cortex during monaural and dichotic listening},
  year      = {2012},
  number    = {1},
  pages     = {78--89},
  volume    = {107},
  publisher = {American Physiological Society Bethesda, MD},
}

@Article{akram2016dynamic,
  author    = {Akram, Sahar and Simon, Jonathan Z. and Babadi, Behtash},
  journal   = {IEEE Trans. on Biomedical Engineering},
  title     = {Dynamic estimation of the auditory temporal response function from {MEG} in competing-speaker environments},
  year      = {2016},
  number    = {8},
  pages     = {1896--1905},
  volume    = {64},
  publisher = {IEEE},
}

@Article{o2015attentional,
  author  = {O'sullivan, James A. and Power, Alan J. and Mesgarani, Nima and Rajaram, Siddharth and Foxe, John J. and Shinn-Cunningham, Barbara G. and Slaney, Malcolm and Shamma, Shihab A. and Lalor, Edmund C.},
  journal = {Cerebral Cortex},
  title   = {Attentional selection in a cocktail party environment can be decoded from single-trial {EEG}},
  year    = {2015},
  number  = {7},
  pages   = {1697--1706},
  volume  = {25},
}

@Article{geirnaert2021electroencephalography,
  author    = {Geirnaert, Simon and Vandecappelle, Servaas and Alickovic, Emina and De Cheveigne, Alain and Lalor, Edmund and Meyer, Bernd T. and Miran, Sina and Francart, Tom and Bertrand, Alexander},
  journal   = {IEEE Signal Processing Magazine},
  title     = {Electroencephalography-based auditory attention decoding: Toward neurosteered hearing devices},
  year      = {2021},
  number    = {4},
  pages     = {89--102},
  volume    = {38},
  publisher = {IEEE},
}

@Article{ding2012emergence,
  author    = {Ding, Nai and Simon, Jonathan Z.},
  journal   = {Proc. of the National Academy of Sciences},
  title     = {Emergence of neural encoding of auditory objects while listening to competing speakers},
  year      = {2012},
  number    = {29},
  pages     = {11854--11859},
  volume    = {109},
  publisher = {National Academy of Sciences},
}

@Article{Wang2023TASA,
  author  = {Wang, Ziwei and Zhang, Wen and Li, Siyang and Chen, Xinru and Wu, Dongrui},
  journal = {Journal of Neural Engineering},
  title   = {Unsupervised domain adaptation for cross-patient seizure classification},
  year    = {2023},
  number  = {6},
  pages   = {066002},
  volume  = {20},
}

@Article{wang2025cst,
  author    = {Wang, Ziwei and Li, Siyang and Wu, Dongrui},
  journal   = {National Science Review},
  title     = {Canine {EEG} helps human: Cross-species and cross-modality epileptic seizure detection via multi-space alignment},
  year      = {2025},
  number    = {6},
  pages     = {nwaf086},
  volume    = {12},
  publisher = {Oxford University Press},
}

@Article{rotaru2024we,
  author  = {Rotaru, Iustina and Geirnaert, Simon and Heintz, Nicolas and Van de Ryck, Iris and Bertrand, Alexander and Francart, Tom},
  journal = {Journal of Neural Engineering},
  title   = {What are we really decoding? Unveiling biases in {EEG}-based decoding of the spatial focus of auditory attention},
  year    = {2024},
  number  = {1},
  pages   = {016017},
  volume  = {21},
}

@Article{kuruvila2021inference,
  author    = {Kuruvila, Ivine and Demir, Kubilay Can and Fischer, Eghart and Hoppe, Ulrich},
  journal   = {IEEE Trans. on Biomedical Engineering},
  title     = {Inference of the selective auditory attention using sequential {LMMSE} estimation},
  year      = {2021},
  number    = {12},
  pages     = {3501--3512},
  volume    = {68},
  publisher = {IEEE},
}

@InProceedings{fu2021auditory,
  author    = {Fu, Zhen and Wang, Bo and Wu, Xihong and Chen, Jing},
  booktitle = {European Signal Processing Conf.},
  title     = {Auditory attention decoding from {EEG} using convolutional recurrent neural network},
  year      = {2021},
  address   = {virtual},
  month     = {Aug.},
  pages     = {970--974},
}

@Article{Marzie2017,
  author  = {Marzieh, Haghighi and Mohammad, Moghadamfalahi and Murat, Akcakaya and Barbara, G. Shinn-Cunningham and Deniz, Erdogmus},
  journal = {IEEE Trans. on Neural Systems and Rehabilitation Engineering},
  title   = {A graphical model for online auditory scene modulation using {EEG} evidence for attention},
  year    = {2017},
  number  = {11},
  pages   = {1970--1977},
  volume  = {25},
}

\end{document}